\documentclass[11pt, a4paper]{article}
\pdfoutput=1

\usepackage{amsmath,amssymb}
\usepackage{epsfig}
\usepackage{hyperref}
\usepackage{color}
\usepackage{cite}
\usepackage{multirow}

\definecolor{myred}{rgb}{0.7, 0, 0}
\definecolor{myblue}{rgb}{0, 0, 0.7}
\definecolor{mygreen}{rgb}{0.04, 0.7, 0.5}

\hypersetup{colorlinks,citecolor=myred,linkcolor=myblue,urlcolor=myblue,linktocpage=true}

\newcommand{\hhref}[1]{\href{http://arxiv.org/abs/#1}{arXiv:#1}}

\newcommand{\be}{\begin{equation}}
\newcommand{\ee}{\end{equation}}
\newcommand{\bea}{\begin{eqnarray}}
\newcommand{\eea}{\end{eqnarray}}
\newcommand{\eq}[1]{Eq.~(\ref{#1})}

\newcommand{\LIR}{\Lambda_{\rm IR}}
\newcommand{\gIR}{g_{*}}

\numberwithin{equation}{section}

\title{
\vspace{-2cm}
\begin{flushright}
\small{CERN-TH-2016-065}
\end{flushright}
\vspace{3cm}
\bf \LARGE
Flavor hierarchies from dynamical scales
\vspace{.2cm}}
\date{}
\author{
{\large Giuliano Panico$^{a}$, Alex Pomarol$^{a,b,c}$}\\
[10mm]
\normalsize\itshape $^a$ IFAE and BIST, Universitat Aut\`onoma de Barcelona, 08193~Bellaterra,~Barcelona\\
\normalsize\itshape $^b$ CERN, Theory Division, Geneva, Switzerland\\
\normalsize\itshape $^c$ Dept.~de~F\'isica, Universitat Aut{\`o}noma de Barcelona, 08193~Bellaterra,~Barcelona\\
}

\begin{document}
\maketitle
\begin{abstract}
\medskip
\noindent
One  main obstacle  for any beyond the SM (BSM) scenario solving the hierarchy problem
is  its  potentially  large contributions to electric dipole moments. 
An elegant way to avoid this  problem is  to have   the  light  SM fermions  couple to  the BSM  sector
 only through  bilinears, $\bar ff$.
 This possibility  can be neatly implemented  in  composite Higgs models.
We study the implications of dynamically generating 
the fermion Yukawa couplings   at different scales,
relating   larger scales to  lighter  SM fermions. 
We show that  all flavor and CP-violating constraints  can be easily accommodated
for  a BSM scale of few TeV, without requiring  any extra symmetry.
Contributions to  $B$ physics are mainly mediated by the top, giving 
a predictive    pattern   of deviations in  $\Delta F=2$ and   $\Delta F=1$ flavor observables
 that could be seen in future experiments.
\end{abstract}

\newpage


\section{Introduction}

An attractive  solution to the hierarchy problem is to require that the Higgs is not an elementary particle, but a composite
 state arising from some strongly-coupled sector at TeV energies.
This possibility has   important implications for the theory of flavor.
Contrary to models with an elementary Higgs in which  the structure of Yukawa couplings  
 can have its origin  at very high energies,  as large as   the Planck scale,
in  composite Higgs models the origin of flavor must be addressed  at  much lower energies.
This is because the Higgs is associated with 
a composite  operator  of the strong sector  ${\cal O}_H$ whose dimension $d_H$ must be larger than
one to avoid the hierarchy problem,\footnote{For the hierarchy problem
what is in fact needed is that the dimension of 
the gauge-singlet term $\mathcal{O}_H \mathcal{O}_H^\dagger$ is larger than $\sim 4$, to avoid relevant operators  in the theory. In strongly-coupled theories with  a  large-$N$ expansion this  implies $d_H\geq 2$, 
but this is not true  in general. Nevertheless, bounds from conformal bootstrap \cite{Rattazzi:2008pe}
 indicate that  it is not possible to have $d_H\sim 1$
together  with  Dim$[\mathcal{O}_H \mathcal{O}_H^\dagger]\gtrsim 4$. Being conservative,  we will be considering here $d_H\gtrsim 2$.}  
implying  that  $\bar f_L {\cal O}_Hf_R$  has dimension larger than 4, that is
to say that the Yukawa   couplings   are irrelevant at low energies.
Therefore, if  $\bar f_L{\cal O}_H f_R $  are   generated at very high energies, e.g.~the Planck scale,
 fermion masses  will  be too small  at the electroweak scale.

Different approaches to flavor in composite Higgs models
have been considered. The most popular one is partial compositeness,
in which the SM fermions  $f_i$ get masses by mixing  linearly with an operator of the  strong sector:
\be
{\cal L}_{\rm lin}=\epsilon_{f_i}\,  \bar f_i\, {\cal O}_{f_i}\, .
\label{linearmix}
\ee
At the strong scale $\LIR\sim$ TeV, which determines the mass-gap of the model,
and   at which the Higgs emerges as a composite state,
the  fermion Yukawa couplings are generated  with a  pattern
\be
{\cal Y}_{f}\sim \gIR \epsilon_{f_i}\epsilon_{f_j}\,,
\label{yuka}
\ee
where $1< \gIR\lesssim 4\pi$  characterizes  the  coupling  in the strong sector. 
The appealing feature of these scenarios, usually called ``anarchic partial compositeness''~\cite{anarchic},  is the fact that the 
smallness of the mixing $\epsilon_{f_i}$ can simultaneously explain the smallness of the fermion masses and mixing angles.
Nevertheless, this approach also predicts flavor-violating higher-dimensional operators of order \cite{Agashe:2004cp}
\be
\frac{\gIR^2}{16\pi^2}\frac{\gIR v}{\LIR^2}\epsilon_{f_i}\epsilon_{f_j}\, \bar f_i \sigma_{\mu\nu}  f_j\,  g F^{\mu\nu}\   ,\ \ \  \ \
\frac{\gIR^2}{\LIR^2}\epsilon_{f_i}\epsilon_{f_j}\epsilon_{f_k}\epsilon_{f_l}\,  \bar f_i\gamma^\mu  f_j  \bar f_k \gamma_\mu f_l\,,
\label{fcnc}
\ee
where $v\simeq 174$ GeV.  The operators in~\eq{fcnc} lead for $\LIR\sim$ TeV to large contributions to the  electron
and neutron electric dipole moment (EDM),  
 $\mu\to e\gamma$ and $\epsilon_K$, above the experimental bounds \cite{KerenZur:2012fr}
 (see also Refs.~\cite{anarchic_bounds,Bauer:2009cf,Konig:2014iqa,Panico:2015jxa}), as shown in Table~\ref{tab:bounds2}.
 Taking $\LIR$ above the TeV  is possible, but at the price of fine-tuning  the electroweak scale.\footnote{Alternative constructions  have  been recently   proposed 
based on  composite Twin Higgs in which  the scale of compositeness can be pushed up without introducing
additional tuning in the Higgs potential~\cite{Csaki:2015gfd}. It is also possible to reduce some bounds by taking smaller $\gIR$, but this implies reducing the UV cutoff (see for example Ref.~\cite{Croon:2015wba}).}

An interesting alternative to the above  approach  is to consider  the right-handed quarks to 
be fully composite \cite{Redi:2011zi}.
If the strong sector has an  accidental $SU(3)$ flavor symmetry and CP symmetry (something not difficult to envisage as it occurs in QCD),
the flavor bounds can be easily satisfied. 
Indeed, in this case the whole  flavor structure comes only  from the linear mixing of the left-handed fermions  with  the strong sector that must   then  be proportional to the SM Yukawas ${\cal Y}_f$, as in models with minimal flavor violation (MFV) \cite{Isidori:2015oea}.
Therefore  flavor bounds are easily satisfied for $\LIR\sim$ TeV.
 Nevertheless, due to the compositeness of the 
right-handed  quarks,   $4$-fermion contact interactions,  as for example, 
\be
\frac{g_*^2}{\LIR^2} (\bar u_R\gamma_\mu u_R)^2\,,
\ee
 lead to large deviation in dijets distributions, $pp\to jj$, at high energies,
and sizable production cross sections for composite resonances in the multi-TeV mass
 range are predicted~\cite{Domenech:2012ai,Redi:2013eaa,Delaunay:2013pwa}.
All these effects have not been observed at the LHC and severely constrain these models.
Similar  results can be found in variations of these ideas
 with other composite SM fermions \cite{Flavor_symm_models}.

Wrapping up, composite Higgs models must address the SM flavor structure at low energies,
giving  then     unequivocal  predictions for flavor observables.
The models proposed so far seem to clash with some  experimental data.
Although  extra flavor and CP symmetries could be imposed, for example in  the mixing terms $\epsilon_{f_i}$,
to avoid certain experimental bounds, 
it is unclear how these symmetries could emerge in the model. One  needs to specify the dynamics of the model
to understand whether flavor and CP symmetries can arise  accidentally at low energies.

Here we would like to  put forward    a   deviation from the  anarchic  paradigm
that can avoid these severe flavor and CP-violating constraints. 
The idea is to assume that  the operators ${\cal O}_{f_i}$ of \eq{linearmix},  that mediate the mixing between the SM fermions
and the Higgs, get an effective   mass  at   some  energy scale $\Lambda_{f_i}\gg\LIR\sim$ TeV,
and then decouple from the strong sector.
This   implies  that    Yukawa-like couplings
\be
{\cal L}_{\rm bil}\sim\bar f_i{\cal O}_{H}f_j\,,
\label{bilinearmix}
\ee
are generated  at scales larger than   $\LIR$, 
avoiding in this way  sizable  contributions to  flavor and CP-violating observables.
The  hierarchies in the  fermion spectrum  of the SM and the small flavor mixing angles 
could be now explained by  the different  scales $\Lambda_{f_i}$ instead of the small $\epsilon_{f_i}$.
The larger the $\Lambda_{f_i}$, the smaller the Yukawa coupling for $f_i$.
Without  imposing  any  extra symmetry in the model, we will derive 
by  simple power-counting which are the strongest flavor and CP-violating constraints,
independently   of the details of the  models.
We find that  top-mediated processes give the largest  contribution to flavor-violating observables.
These   are characterized by only two operators.
One operator generates the  $\Delta F=2$ processes    $\epsilon_K$,  $\Delta M_{B_d}$  and $\Delta M_{B_s}$  
at a level   close to the present experimental constraints for $\LIR\sim$ {\it few} TeV.
 The second operator  leads to  flavor-violating  $Z$-couplings, 
contributing simultaneously to   $K\to \mu^+\mu^-$, $\epsilon'/\epsilon$, $B\to (X)\ell\ell$
and  $Z\to b\bar b$ with  a size also close to the experimental bounds.
There are  also important contributions arising from the scale at which  the charm and strange masses are  generated, 
$10^{7}-10^8$ GeV,  leading  also  to sizable effects to $\epsilon_K$,  and forcing    $d_H\lesssim 2$. 
Contributions to the neutron  EDM are dominated  by the top EDM, 
being  not far from the present experimental  bound.
On the other hand,  in the lepton sector we find that  the dominant contribution to the electron EDM comes at the two-loop level 
from Barr-Zee type diagrams \cite{Barr:1990vd},  and is around the experimental bound,
while   $\mu\to e\gamma$ is  found to be  very  small.
Therefore these scenarios provide  realistic examples  where the  flavor and hierarchy problem 
can be dynamically  solved without contradicting the present experimental data,      and   which
 near future experiments   could be able to explore.
Having proposed a different origin for   fermion masses,
we also analyze  the expected deviations in Higgs couplings.

Our approach to the  small fermion masses is a   reminiscent of the old Extended-Technicolor idea \cite{Dimopoulos:1979es},
in which  masses from \eq{bilinearmix} were generated from an extended gauge sector, or
from integrating heavy fermions \cite{Berezhiani:1992pj}.
Earlier attempts along these lines  were considered recently  in Refs.~\cite{Matsedonskyi:2014iha}
for composite Higgs models.
In these  models, however,  Yukawa-like couplings 
were generated at a single energy scale, and the light quark families were connected
by potentially large mixing angles. 
This leads to additional sizable new-physics effects and to bounds typically
more stringent than the ones we find here. 
Furthermore, the lepton sector,  where the  experimental bounds are the most difficult to satisfy,
 was not considered.

We would like to  close this section  by 
stressing that in  most  scenarios beyond the SM (BSM) that address the hierarchy problem,  including supersymmetry,  
one generically  finds large EDMs.
This is because  fermions have linear couplings to BSM fields. For example, in supersymmetric models fermions
 couple linearly to sfermions and gauginos, leading   generically to sizable EDMs  at the one-loop level.
Therefore, unless  {\it ad hoc} symmetries are imposed  in the  BSM sector,
the only way to avoid these large contributions is to  restrict the SM fermions to have bilinear couplings to the BSM states,
as the scenarios proposed here.
In this case the dominant contributions to EDMs arise at the two-loop level (see diagram Fig.~\ref{fig:bz})
that can be accommodated just below the experimental constraint.

\begin{figure}[t]
\centering
\includegraphics[width=0.45\textwidth]{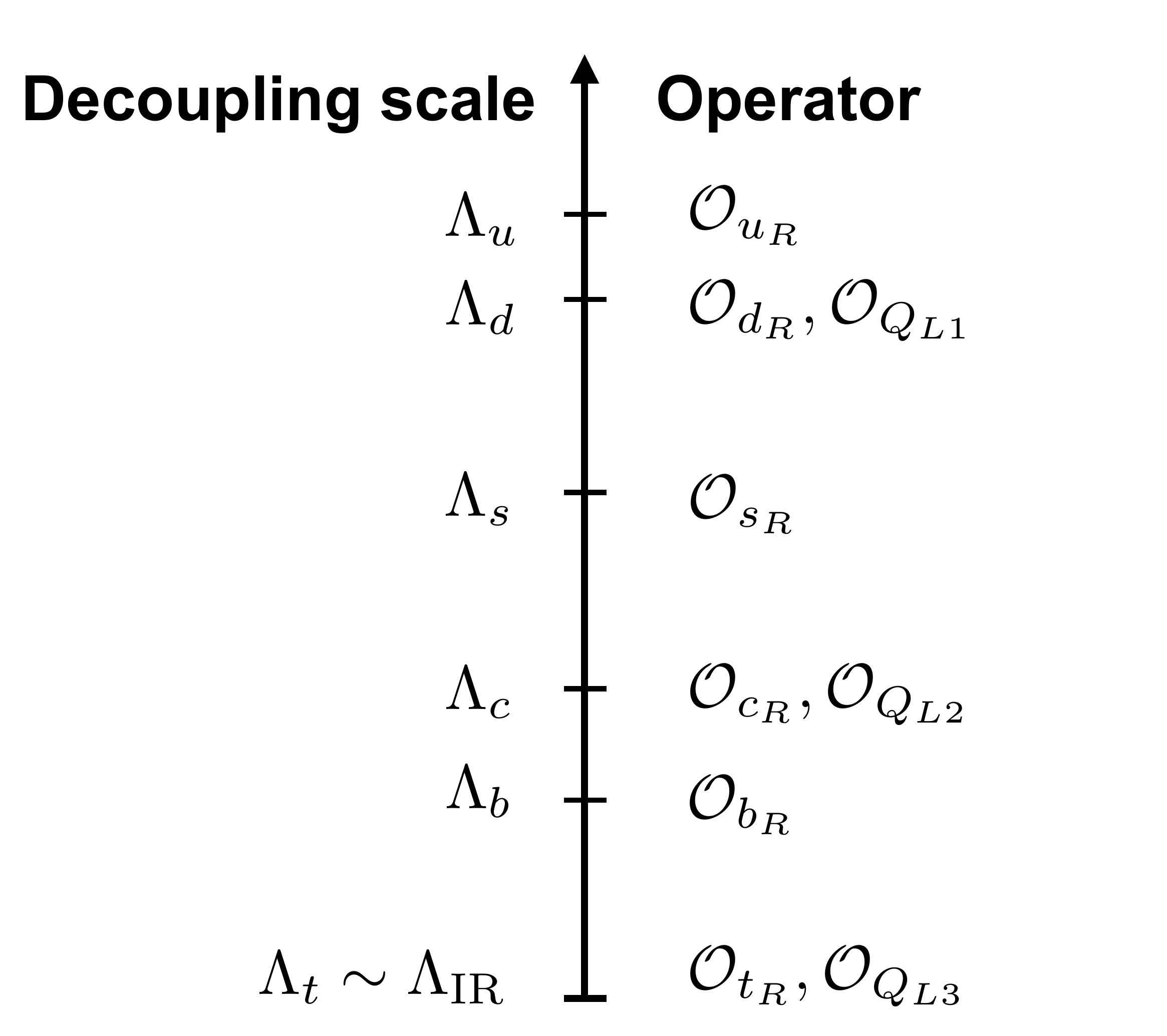}
\caption{\it Energy scale at which the fermionic operators ${\cal O}_{f_i}$ decouple from the strong sector.}
\label{decouplingscales}
\end{figure}

\section{Multiple flavor scales in composite Higgs models}\label{sec:scenarios}

Our  framework  for flavor shares many features of   previous composite Higgs models with  partly-composite fermions via
\eq{linearmix}.
The main  crucial  difference  is  the assumption  that  the operators  ${\cal O}_{f_i}$, 
which are  the portals  of the  SM fermions to the strong sector,
  decouple  at  some scale   $\Lambda_{f_i}$,
generating 
the  Yukawa terms   $\bar f_L{\cal O}_H f_R$ at that scale instead of at $\LIR$ as in the anarchic case.
The decoupling of the operator ${\cal O}_{f_i}$  can be due to the fact that 
 some of the constituents  of  ${\cal O}_{f_i}$ get   a mass $\sim\Lambda_{f_i}$, 
or that a dynamically generated  mass-gap makes  
heavy all composite states created by  ${\cal O}_{f_i}$ (those $|\Psi\rangle$
with $\langle 0|{\cal O}_{f_i}|\Psi\rangle\not=0$).
 Using the AdS/CFT correspondence, we can  easily
visualize   this type of  scenarios by   warped extra-dimensional models  with several branes,
as  the example shown in Fig.~\ref{ads} of  Appendix~\ref{adscft}. 
In what follows we will  estimate the flavor  structure   of  these scenarios
without  restricting     to any specific UV realization.

\begin{figure}[t]
\centering
\includegraphics[width=0.7\textwidth]{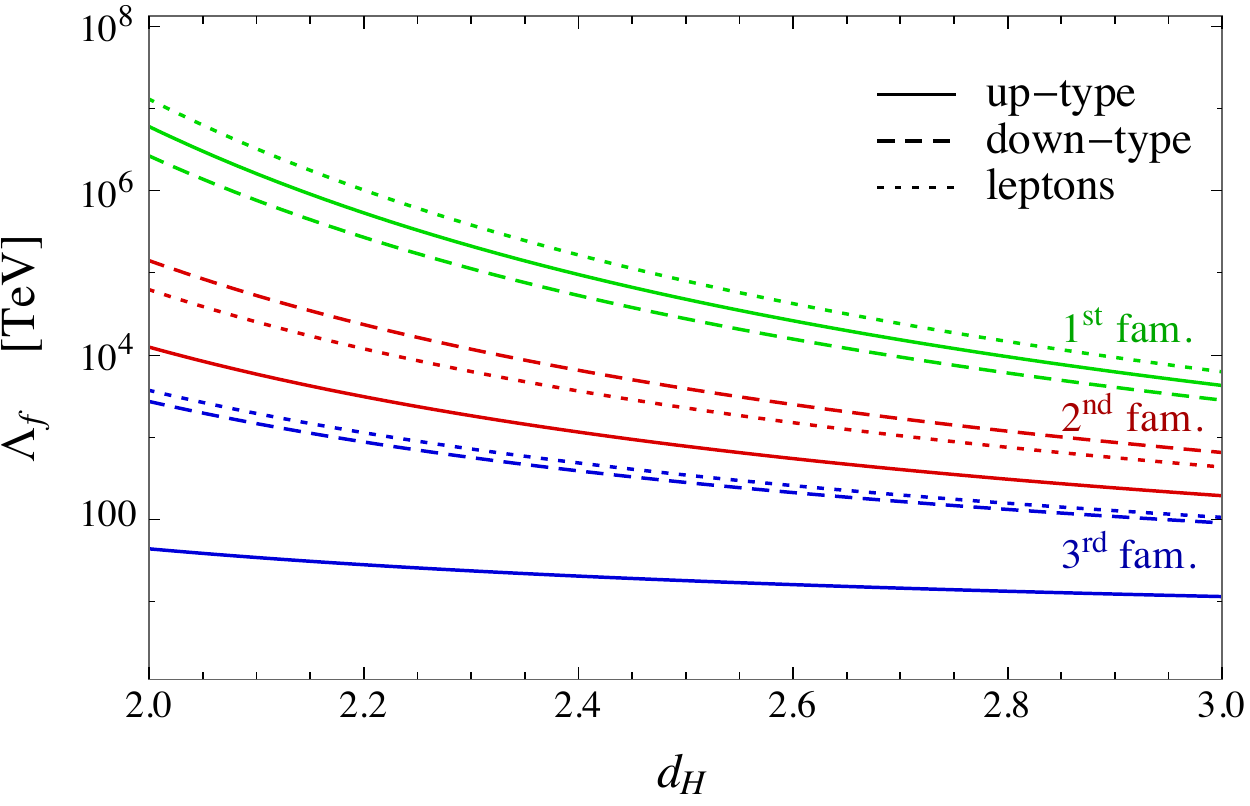}
\caption{\it Upper bound on the scale $\Lambda_f$  (for $f=e,u,d,s,\mu,c,\tau,b,t$ from top to down)
at which  the fermion Yukawas can originate from a bilinear term (\eq{yukawas} with $\epsilon^{(i)}_{f_{Li,Ri}}\sim 1$, $\gIR\sim 4\pi$ and for $\LIR=3$ TeV)   as a function of  $d_H$, the dimension of the Higgs composite operator ${\cal O}_H$. To derive the numerical results we identified the fermion masses with the running masses at  1 TeV \cite{KerenZur:2012fr},
neglecting the effect of running  $m_{f}$ from TeV to $\Lambda_{f}$.
}
\label{lambdaf}
\end{figure}

The scale at which the Yukawa coupling for the SM fermion  $f=u,d,e,...$
is generated is determined by the scale $\Lambda_f$ 
at which  either  ${\cal O}_{f_R}$ or ${\cal O}_{f_L}$ decouple from the strong sector.
We choose these scales following  Fig.~\ref{decouplingscales}.
This is our dynamical assumption. No further symmetries will be imposed.
Other options could also be possible, and we will consider later more economical models with fewer  scales $\Lambda_f$. 
Under the assumption of Fig.~\ref{decouplingscales},  the Yukawa   structure  will be the following. 
Let us  consider first the down-type quark sector.
At the lowest scale   $\Lambda_{b}$, we have 
only one pair of operators ${\cal O}_{Q_{L3}}$ and  ${\cal O}_{b_R}$,
to which only one linear combination of SM left-handed and right-handed  quarks can respectively mix with.
We name these   linear combinations the 3rd family left-handed quark,  $Q_{L3}$, and  right-handed bottom, $b_R$:
\be
{\cal L}^{(3)}_{\rm lin}=\epsilon^{(3)}_{b_{L}}
\bar Q_{L3}\, {\cal O}_{Q_{L3}}+
\epsilon^{(3)}_{b_{R}}
\bar b_R\,  {\cal O}_{b_R}\,.
\ee
Below $\Lambda_b$, after  integrating out ${\cal O}_{b_R}$,  the following Yukawa-like operator is expected to be generated
\begin{equation}
\label{bilinearmix3}
{\cal L}^{(3)}_{\rm bil}= \frac{1}{\Lambda_{b}^{d_H - 1}}   (\epsilon^{(3)}_{b_L} \bar Q_{L3})\mathcal{O}_H ( \epsilon^{(3)}_{b_R}  b_R)\,,
\end{equation}
where  $\mathcal{O}_H$ corresponds to the  lowest-dimensional operator 
that at $\LIR$ projects into the Higgs, $\langle 0|\mathcal{O}_H|H\rangle\not =0$,
and   $d_H$ is its  energy dimension.
At a larger  scale $\Lambda_{s}\gg \Lambda_b$,    we have  another pair of operators
${\cal O}_{Q_{L2}}$ and  ${\cal O}_{s_R}$ present, coupled 
to a  different linear combination of SM fermions.
 By an  $SU(3)$ rotation that does not affect \eq{bilinearmix3}
 we can always go to  the basis where this
linear combination  contains only two quarks, $Q_{L3}$ and  $Q_{L2}$ (this latter is identified with the second family left-handed quark), and similarly for the right-handed sector, $b_R$ and  $s_R$: 
\be
{\cal L}^{(2)}_{\rm lin}=(\epsilon^{(2)}_{b_L} \bar Q_{L3} +\epsilon^{(2)}_{s_L} \bar Q_{L2}) 
\, {\cal O}_{Q_{L2}}+
(\epsilon^{(2)}_{b_R}  b_R+\epsilon^{(2)}_{s_R}  s_R)
\,  {\cal O}_{s_R}\,,
\label{bilinearmix12}
\ee
that below   $\Lambda_s$, after integrating  ${\cal O}_{s_R}$,  leads to 
\begin{equation}
\label{bilinearmix2}
{\cal L}^{(2)}_{\rm bil}= \frac{1}{\Lambda_{s}^{d_H - 1}}   (\epsilon^{(2)}_{b_L} \bar Q_{L3} +\epsilon^{(2)}_{s_L} \bar Q_{L2} )\mathcal{O}_H ( \epsilon^{(2)}_{b_R}  b_R+\epsilon^{(2)}_{s_R}  s_R)\,.
\end{equation}
Finally, at $\Lambda_{d}$, 
after integrating     ${\cal O}_{Q_{L1}}$ and  ${\cal O}_{d_R}$, 
we expect the most general form 
\begin{equation}
{\cal L}^{(1)}_{\rm bil}= \frac{1}{\Lambda_{d}^{d_H - 1}}   
(\epsilon^{(1)}_{b_L} \bar Q_{L3} +\epsilon^{(1)}_{s_L} \bar Q_{L2}+\epsilon^{(1)}_{d_L} \bar Q_{L1} )
\mathcal{O}_H 
( \epsilon^{(1)}_{b_R}  b_R+\epsilon^{(1)}_{s_R}  s_R+\epsilon^{(1)}_{d_R}  d_R)\,.
\label{bilinearmix1}
\end{equation}
Now, at  $\LIR$  we   identify the matrix elements of ${\cal O}_H$ with those of the  SM Higgs $H$, 
which implies the replacement
\footnote{For simplicity we are assuming a single coupling $\gIR$, but in principle the couplings  at the scales $\Lambda_{f}$ could be different.}
\be
{\cal O}_H\rightarrow \gIR \LIR^{d_H-1} H\,,
\ee
in  \eq{bilinearmix3}, \eq{bilinearmix2} and \eq{bilinearmix1}.
Then, for the down sector, we have the following ``onion'' Yukawa structure
\bea
{\cal Y}_{\rm down}&=&
\gIR
 \left(
\begin{array}{ccccc}
\epsilon^{(1)}_{d_L} \epsilon^{(1)}_{d_R}  && \epsilon^{(1)}_{d_L} \epsilon^{(1)}_{s_R}  && \epsilon^{(1)}_{d_L} \epsilon^{(1)}_{b_R} \\
\rule{0pt}{1.25em}
\epsilon^{(1)}_{s_L} \epsilon^{(1)}_{d_R}  &&&&\\
\rule{0pt}{1.25em}
 \epsilon^{(1)}_{b_L} \epsilon^{(1)}_{d_R} &&&&
\end{array}
\right)\left(\frac{\LIR}{\Lambda_d}\right)^{d_H-1}
+\,
\gIR
\left(
\begin{array}{ccccc}
0&& 0 && 0\\
\rule{0pt}{1.25em}
0 && \epsilon^{(2)}_{s_L} \epsilon^{(2)}_{s_R} &&\epsilon^{(2)}_{s_L} \epsilon^{(2)}_{b_R}\\
\rule{0pt}{1.25em}
0 &&\epsilon^{(2)}_{b_L} \epsilon^{(2)}_{s_R}&&
\end{array}
\right)\left(\frac{\LIR}{\Lambda_s}\right)^{d_H-1}\nonumber\\
&&\qquad+\, \gIR
\left(
\begin{array}{ccccc}
0&& 0 && 0\\
\rule{0pt}{1.25em}
0 &&0 &&0\\
\rule{0pt}{1.25em}
0 &&0&& \epsilon^{(3)}_{b_L} \epsilon^{(3)}_{b_R}
\end{array}
\right)\left(\frac{\LIR}{\Lambda_b}\right)^{d_H-1}\,,
\label{yukawad}
\eea
where the entries that are not shown are terms that can be neglected in the limit
in which we take $\Lambda_d\gg \Lambda_s\gg \Lambda_b$.
\eq{yukawad}  leads to the approximate down Yukawa matrix
\be
{\cal Y}_{\rm down}\simeq
\left(
\begin{array}{ccccc}
\ \ \  Y_d && \alpha_R^{ds} Y_d && \alpha_R^{db} Y_d\\
\rule{0pt}{1.25em}
 \alpha_L^{ds} Y_d &&  \ \ \ Y_s && \alpha_R^{sb} Y_s
\\
\rule{0pt}{1.25em}
 \alpha_L^{db} Y_d &&   \alpha_L^{sb} Y_s && \ \ \ Y_b
\end{array}
\right)\,,
\label{matrixYd}
\ee
where 
\be
Y_{f}\equiv \gIR \epsilon^{(i)}_{f_{Li}}\epsilon^{(i)}_{f_{Ri}} \left(\frac{\LIR}{\Lambda_{f}}\right)^{d_H-1}\,,
\label{yukawas}
\ee
are approximately  the SM Yukawas $Y_{f}\simeq m_{f}/v$.
The   $\alpha_{L}$ and $\alpha_R$ in \eq{matrixYd} are ratios of epsilons:
\be
\alpha_{L}^{ds}\sim {\epsilon^{(1)}_{s_L}}/{\epsilon^{(1)}_{d_L}}\ ,\qquad\qquad
\alpha_{L}^{db}\sim {\epsilon^{(1)}_{b_L}}/{\epsilon^{(1)}_{d_L}}\ ,\qquad\qquad
\alpha_{L}^{sb}\sim {\epsilon^{(2)}_{b_L}}/{\epsilon^{(2)}_{s_L}}\,,
\ee
where    $L\rightarrow R$ gives us the $\alpha_R$.
Taking the largest values $\epsilon^{(i)}_{f_{Li,Ri}} \sim 1$ and $\gIR\sim 4\pi$, 
we can obtain from~\eq{yukawas} the  largest values of   $\Lambda_{f}$ that allow to reproduce the SM fermion masses  as a function of  $d_H$,  that we show  in Fig.~\ref{lambdaf}.
For the particular case $d_H=2$, we have
  \be
  \Lambda_{f}\sim  \frac{\gIR}{Y_{f}}\LIR\, ,
  \ee
 that,  for $\LIR\sim 3$ TeV and $\gIR\sim 4\pi$, gives
\be
\Lambda_d\sim 3\times 10^9\ {\rm GeV}\ ,\qquad
\Lambda_s\sim 10^8\ {\rm GeV}\ ,\qquad
\Lambda_b\sim 3\times 10^6\ {\rm GeV}\,.
\label{scaleval}
\ee
\eq{matrixYd}  can be diagonalized by unitary matrices whose structure is approximately
\begin{equation}
V^{\rm down}_L \sim \left(
\begin{array}{ccccc}
1 && \alpha_R^{ds} \frac{Y_d}{Y_s} && \alpha_R^{db} \frac{Y_d}{Y_b}\\
&& 1 &&  \alpha_R^{sb} \frac{Y_s}{Y_b} \\
&& && 1
\end{array}
\right)\,,
\qquad
V^{\rm down}_R \sim \left(
\begin{array}{ccccc}
1 &&  \alpha_L^{ds} \frac{Y_d}{Y_s} &&  \alpha_L^{db} \frac{Y_d}{Y_b}\\
&& 1 &&  \alpha_L^{sb} \frac{Y_s}{Y_b} \\
&& && 1
\end{array}
\right)\,,
\label{rotationLR}
\end{equation}
where we omit some  $ij$-entries as they are  of similar size as  their transpose $ji$-entries.

We can  proceed  in a similar way for the up sector.
The large Yukawa coupling of the top implies that   this must arise at $\LIR$  as in the anarchic case,
so we associate $\Lambda_t\sim\LIR$.
The  Yukawa  matrix is expected to have the structure
\be
{\cal Y}_{\rm up}\simeq
\left(
\begin{array}{ccccc}
\ \ \  Y_u && \alpha_R^{uc} Y_u && \alpha_R^{ut} Y_u\\
\rule{0pt}{1.25em}
 \alpha_L^{uc} Y_u &&  \ \ \ Y_c && \alpha_R^{ct} Y_c
\\
\rule{0pt}{1.25em}
 \alpha_L^{ut} Y_u &&   \alpha_L^{ct} Y_c && \ \ \ Y_t
\end{array}
\right)\,.
\label{matrixYu}
\ee
We must point out however that there can be extra contributions coming from $\Lambda_{d,s,b}$.
The most important ones come from $\Lambda_{d}$ 
where it is possible  to generate
\begin{equation}
\Delta {\cal L}^{(1)}_{\rm bil}= \frac{1}{\Lambda_{d}^{d_H - 1}}   
\epsilon^{(1)}_{d_L} \bar Q_{L1} 
\tilde {\cal O}_H 
( \tilde \epsilon^{(1)}_{t_R}  t_R+\tilde \epsilon^{(1)}_{c_R}  c_R)\,,
\label{bilinearmix1extra}
\end{equation}
that leads to contributions  to  the  entries  $({\cal Y}_{\rm up})_{13}\sim({\cal Y}_{\rm up})_{12}\sim  Y_d$ 
that can be slightly larger than those
in \eq{matrixYu} since $Y_d > Y_u$. 
We  absorb these contributions in \eq{matrixYu} by a redefinition of $\alpha_R^{uc,ut}$. 
Similarly, 
${\cal Y}_{\rm down}$ can receive extra  contributions from $\Lambda_{u,c,t}$. The largest expected 
one is from $\Lambda_{c}$ where we can have
\be
\frac{1}{\Lambda_{c}^{d_H - 1}}   
 \bar Q_{L2} \mathcal{O}_H      b_R\,,
 \label{bilinearmix1extra2}
 \ee
that leads to  $({\cal Y}_{\rm down})_{23}\sim  Y_c$ that is parametrically a factor
$Y_c/Y_s\sim 10$ larger than the corresponding entry  in \eq{matrixYd}. Again, we  
 absorb this contribution   in a redefinition of $\alpha_R^{sb}$. 
We must add however that if  the strong sector had an $SU(3)$ flavor symmetry, the contributions in \eq{bilinearmix1extra}
and \eq{bilinearmix1extra2} would be zero, as they originate  from the off-diagonal  interactions in the strong sector,
 ${\cal O}_{Q_{L1}}\tilde {\cal O}_H {\cal O}_{t_R,c_R}$ and  ${\cal O}_{Q_{L2}}{\cal O}_H{\cal O}_{b_R}$
respectively.

Since  the mass hierarchies in the up sector are larger than in the down sector, 
we have that  the CKM matrix $V_{\textrm{CKM}}$ is mainly dominated by the down rotation:
\be 
V_{\textrm{CKM}} \sim  (V^{\rm down}_L)^\dagger\,,
\label{vckm}
\ee
impliying the following conditions on the $\alpha_L$'s of the down-Yukawa matrix:
 \begin{equation}
\alpha_R^{ds} \frac{m_d}{m_s} \simeq (V_{\textrm{CKM}})_{21} \simeq \lambda_c\ ,
\ \ \ \
\alpha_R^{sb} \frac{m_s}{m_b} \simeq (V_{\textrm{CKM}})_{32}\simeq \lambda_c^2\ ,
\ \ \ \
\alpha_R^{db} \frac{m_d}{m_b} \simeq (V_{\textrm{CKM}})_{31} \simeq \lambda_c^3\,,
\label{alphas}
\end{equation}
where $\lambda_c \simeq 0.22$ is the  Cabibbo angle. From the estimate
\be
\frac{m_d}{m_s} \sim  \frac{m_s}{m_b} \sim \lambda_c^2\,,
\ee
we obtain using \eq{alphas}  that   $\alpha_R^{ds,db}$  must be slightly larger than one, in particular,
\be
\alpha_R^{ds} \sim \alpha_R^{db} \sim 1/\lambda_c\ ,\qquad \alpha_R^{sb}\sim 1\, .
\label{alphar}
\ee
This can be easily accommodated by having   $\epsilon^{(1)}_{s_R,b_R}$   slightly smaller than one
(and a suppression of \eq{bilinearmix1extra2}).
On the other hand, the $\alpha_L$ are not constrained at all by the CKM angles,
and could even be very small  if some  mixings are zero. For example, this could be the case if
$\epsilon^{(1)}_{s_L,b_L}\approx 0$  due to  some accidental discrete symmetry  at $\Lambda_d$, as discussed in Appendix~\ref{app:z2}.
Notice  that in the limit  $\epsilon^{(1)}_{s_L,b_L}\to 0$ the rotation matrix $V_R^{\rm down}$ 
is not anymore given by \eq{rotationLR} but by \eq{brotation}.
Nevertheless, we emphasize that the framework for flavor  proposed here does not need any accidental symmetry
to pass the phenomenological constraints, as we discuss below.

\section{Implications in flavor and CP-violation physics}

At each scale $\Lambda_{f}$ we have potentially new flavor-violating  contributions,
arising   from higher-dimensional operators made of  SM fermions.
We can estimate these effects using power-counting arguments, since no
flavor symmetries are assumed in our scenarios.
The most important  higher-dimensional operators are  $4$-quark operators, that contribute to $\Delta F =2$ transitions,
$2$-quark-$2$-Higgs operators that generate $\Delta F =1$ effects,
and dipole operators contributing to processes such as  $\mu\to e\gamma$  or EDMs. 
We collect the most important experimental bounds in Table~\ref{expbounds}.

\subsection{\boldmath $\Delta F =2$ transitions}

We start considering  $4$-quark operators   arising at the lowest scale $\Lambda_t\sim \LIR$.
These are  operators containing only top components, $Q_{L3}$ and $t_R$,
namely~\footnote{These estimates are valid even if  $\Lambda_t>\LIR$ and the top partners are heavier than
$\LIR$. Nevertheless, for   top partners lighter than $\LIR$, as could be needed in these scenarios to obtain a viable Higgs mass and minimize the amount of tuning~\cite{light_partners,Pomarol:2012qf}, the $4$-fermion operators get enhanced.
For a discussion see Ref.~\cite{Panico:2015jxa}.}
\be
\frac{Y_t^2x_t^2}{\LIR^2}(\overline Q_{L3} \gamma^\mu Q_{L3})^2\ ,\ \ \ 
\frac{Y_t^2}{\LIR^2}(\overline Q_{L3}  t_R)(\overline t_R Q_{L3}   )\ ,\ \ \  
\frac{Y_t^2/x_t^2}{\LIR^2}(\overline t_R \gamma^\mu t_R)^2\,,
\label{4top}
\ee
where we defined $x_t=\epsilon^{(3)}_{t_L}/\epsilon^{(3)}_{t_R}$.

\begin{table}
\footnotesize
\centering
\begin{tabular}{@{\;}c@{\;}|@{\;}c@{\;}|c|c|@{\;}c@{\;}}
Observable & Operator & Re part & Im part & Reference\\
\hline
\hline
\multirow{3}{*}{\raisebox{-.25em}{$\Delta M_K; \epsilon_K$}}
& \rule{0pt}{1.1em}${\cal Q}_1^{sd} = (\overline s_L \gamma^\mu d_L)^2$ & $1.1 \times 10^3$ & $1.7 \times 10^4$
& \multirow{3}{*}{\cite{Bona:2007vi,Isidori:2015oea}}\\
& \rule{0pt}{1.1em}${\cal Q}_2^{sd} = (\overline s_R d_L)^2$,  $\widetilde {\cal Q}_2^{sd} = (\overline s_L d_R)^2$ & $7.3 \times 10^3$ & $1.2 \times 10^5$ \\
& \rule[-.5em]{0pt}{1.6em}${\cal Q}_4^{sd} = (\overline s_R d_L)(\overline s_L d_R)$ & $1.2 \times 10^4$ & $2.0 \times 10^5$ \\
\hline
\multirow{3}{*}{\raisebox{-.25em}{$\Delta M_{B_d}; S_{\psi K_S}$}}
& \rule{0pt}{1.1em}${\cal Q}_1^{bd} = (\overline b_L \gamma^\mu d_L)^2$ & $6.6 \times 10^2$ & $ 9.5 \times 10^2$
& \multirow{3}{*}{\cite{Bona:2007vi,Isidori:2015oea}}\\
& \rule{0pt}{1.1em}${\cal Q}_2^{bd} = (\overline b_R d_L)^2$,  $\widetilde {\cal Q}_2^{bd} = (\overline b_L d_R)^2$ & $1.2 \times 10^3$ & $1.7 \times 10^3$\\
& \rule[-.5em]{0pt}{1.6em}${\cal Q}_4^{bd} = (\overline b_R d_L)(\overline b_L d_R)$ & $1.6 \times 10^3$ & $2.3 \times 10^3$ \\
\hline
\multirow{3}{*}{\raisebox{-.25em}{$\Delta M_{B_s}; S_{\psi\phi}$}}
& \rule{0pt}{1.1em}${\cal Q}_1^{bs} = (\overline b_L \gamma^\mu s_L)^2$ & $1.4 \times 10^2$ & $2.4 \times 10^2$
& \multirow{3}{*}{\cite{Bona:2007vi,Isidori:2015oea}}\\
& \rule{0pt}{1.1em}${\cal Q}_2^{bs} = (\overline b_R s_L)^2$,  $\widetilde {\cal Q}_2^{bs} = (\overline b_L s_R)^2$ & $1.3 \times 10^2$ & $2.2 \times 10^2$\\
& \rule[-.5em]{0pt}{1.6em}${\cal Q}_4^{bs} = (\overline b_R s_L)(\overline b_L s_R)$ & $3.4 \times 10^2$ & $5.9 \times 10^2$\\
\hline
\multirow{3}{*}{\raisebox{-.25em}{$\Delta M_{D}; |q/p|, \phi_D$}}
& \rule{0pt}{1.1em}${\cal Q}_1^{cu} = (\overline c_L \gamma^\mu u_L)^2$ & $1.3 \times 10^3$ & $3.2 \times 10^3$
& \multirow{3}{*}{\cite{Bona:2007vi,Isidori:2015oea}}\\
& \rule{0pt}{1.1em}${\cal Q}_2^{cu} = (\overline c_R u_L)^2$,  $\widetilde {\cal Q}_2^{cu} = (\overline c_L u_R)^2$  & $2.5 \times 10^3$ & $5.8 \times 10^3$\\
& \rule[-.5em]{0pt}{1.6em}${\cal Q}_4^{cu} = (\overline c_R u_L)(\overline c_L u_R)$ & $4.2 \times 10^3$ & $9.5 \times 10^3$\\
\hline
\rule[-.6em]{0pt}{1.7em}& $(\overline Q_{Li} ({\cal Y}_{\rm up} {\cal Y}_{\rm up}^\dagger)_{ij} \gamma^\mu Q_{Lj})^2$ &
\multicolumn{2}{c|@{\;}}{$5$} & \cite{Bona:2007vi}\\
\hline
\hline
\rule[-.5em]{0pt}{1.6em}$b \rightarrow s \ell^+ \ell^-$ & \rule{0pt}{1.3em}$(\overline s_L \gamma^\mu b_L) H^\dagger i \overleftrightarrow D_\mu H$ & $23$ & $16$ & \cite{Altmannshofer:2012az,Guadagnoli:2013mru}\\
\hline
\parbox{8em}{\centering $K_L \rightarrow \mu^+ \mu^-$,\\ $K^+ \rightarrow \pi^+ \nu \overline \nu, \epsilon'/\epsilon$}
& \rule[-1.05em]{0pt}{2.7em}$(\overline s_L \gamma^\mu d_L) H^\dagger i \overleftrightarrow D_\mu H$ &
\multicolumn{2}{c|@{\;}}{$225$}
& \cite{Bauer:2009cf,Buras:2011ph}\\
\hline
\rule[-.5em]{0pt}{1.6em}$Z \rightarrow b\overline b$ & \rule{0pt}{1.3em}$(\overline b_L \gamma^\mu b_L) H^\dagger i \overleftrightarrow D_\mu H$ &
\multicolumn{2}{c|@{\;}}{$5.5$} & $|\delta g_{b_L}| \lesssim 10^{-3}$~\cite{Grojean:2013qca,Ghosh:2015wiz}\\
\hline
\multirow{4}{*}{\raisebox{-.25em}{$B \rightarrow X_s \gamma$}} & \rule{0pt}{1.0em}$m_b\, \overline s_L \sigma^{\mu\nu} e F_{\mu\nu} b_R$ & $8.9$\hspace{1.25em}$35$ & $18$ & \multirow{4}{*}{\raisebox{-.0em}{\cite{Konig:2014iqa}}}\\
& \rule{0pt}{1.em}$m_b\, \overline s_R \sigma^{\mu\nu} e F_{\mu\nu} b_L$ & $18$ & $16$ & \\
& \rule{0pt}{1.em}$m_b\, \overline s_L \sigma^{\mu\nu} g_s G_{\mu\nu} b_R$ & $4.3$\hspace{1.25em}$17$ & $8.6$ &\\
& \rule[-0.5em]{0pt}{1.5em}$m_b\, \overline s_R \sigma^{\mu\nu} g_s G_{\mu\nu} b_L$ & $8.5$ & $8.5$ & \\
\hline
\multirow{4}{*}{\raisebox{-.25em}{$B \rightarrow X_d \gamma$}} & \rule{0pt}{1.em}$m_b\, \overline d_L \sigma^{\mu\nu} e F_{\mu\nu} b_R$ & $47$\hspace{1.25em}$19$ & $37$\hspace{1.25em}$24$ & \multirow{4}{*}{\raisebox{-.0em}{\cite{Konig:2014iqa}}}\\
& \rule{0pt}{1.em}$m_b\, \overline d_R \sigma^{\mu\nu} e F_{\mu\nu} b_L$ & $30$ & $30$ & \\
& \rule{0pt}{1.em}$m_b\, \overline d_L \sigma^{\mu\nu} g_s G_{\mu\nu} b_R$ & $22$\hspace{1.25em}$9$ & $18$\hspace{1.25em}$12$ & \\
& \rule[-0.5em]{0pt}{1.5em}$m_b\, \overline d_R \sigma^{\mu\nu} g_s G_{\mu\nu} b_L$ & $14$ & $14$ & \\
\hline
$K \rightarrow 2 \pi, \epsilon'/\epsilon$ & \rule[-0.5em]{0pt}{1.6em}$m_s\, \overline s_{L,R} \sigma^{\mu\nu} g_s G_{\mu\nu} d_{R,L}$ & & $35$ & \cite{Konig:2014iqa}\\
\hline
$D \rightarrow KK, \pi\pi$ & \rule[-0.5em]{0pt}{1.6em}$m_c\, \overline c_{L,R} \sigma^{\mu\nu} g_s G_{\mu\nu} u_{R,L}$ & & $27$ & \cite{Konig:2014iqa}\\
\hline
\hline
\multirow{7}{*}{\raisebox{-1.5em}{Neutron EDM}} & \rule{0pt}{1.1em}$m_d\, \overline d_L \sigma^{\mu\nu} e F_{\mu\nu} d_R$ & & $39$ &
\multirow{7}{*}{\raisebox{-1.5em}{$|d_n| < 2.9 \times 10^{-26}\, e\ \mathrm{cm}$~\cite{Agashe:2014kda,Konig:2014iqa,EDMs}}}\\
& \rule{0pt}{1.0em}$m_u\, \overline u_L \sigma^{\mu\nu} e F_{\mu\nu} u_R$ & & $14$ & \\
& \rule{0pt}{1.0em}$m_d\, \overline d_L \sigma^{\mu\nu} g_s G_{\mu\nu} d_R$ & & $48$ & \\
& \rule{0pt}{1.0em}$m_u\, \overline u_L \sigma^{\mu\nu} g_s G_{\mu\nu} u_R$ & & $18$ & \\
& \rule{0pt}{1.0em}$m_c\, \overline c_L \sigma^{\mu\nu} g_s G_{\mu\nu} c_R$ & & $15$ & \\
& \rule{0pt}{1.0em}$m_b\, \overline b_L \sigma^{\mu\nu} g_s G_{\mu\nu} b_R$ & & $8.4$ & \\
& \rule[-0.5em]{0pt}{1.5em}$m_t\, \overline t_L \sigma^{\mu\nu} g_s G_{\mu\nu} t_R$ & & $3.7$ & \\
\hline
\hline
\rule[-0.5em]{0pt}{1.6em} Electron EDM & $m_e\, \overline e_L \sigma^{\mu\nu} e F_{\mu\nu} e_R$ & & $480$ & $|d_e| < 0.87 \times 10^{-28} \,e\ \mathrm{cm}$~\cite{Agashe:2014kda}\\
\hline
\rule{0pt}{1.1em}$\mu \rightarrow e \gamma$ & $m_\mu\, \overline \mu \sigma^{\mu\nu} e F_{\mu\nu} e_{R,L}$ & \multicolumn{2}{c|@{\;}}{$900$} & $BR(\mu \rightarrow e \gamma) < 5.7 \times 10^{-13}$~\cite{Agashe:2014kda}\\
\rule{0pt}{1.1em}$\tau \rightarrow \mu \gamma$ & $ m_\tau\, \overline \tau \sigma^{\mu\nu} e F_{\mu\nu} \mu_{R,L}$ & \multicolumn{2}{c|@{\;}}{$34$} & $BR(\tau \rightarrow \mu \gamma) < 4.4 \times 10^{-8}$~\cite{Agashe:2014kda}\\
\rule{0pt}{1.1em}$\tau \rightarrow e \gamma$ & $m_\tau\, \overline \tau \sigma^{\mu\nu} e F_{\mu\nu} e_{R,L}$ & \multicolumn{2}{c|@{\;}}{$37$} & $BR(\tau \rightarrow e \gamma) < 3.3 \times 10^{-8}$~\cite{Agashe:2014kda}
\end{tabular}
\caption{\it Experimental bounds on  new physics contributions to  flavor and CP-violating  operators. The bounds are
computed at an energy scale $\mu = 1$ TeV and are expressed as constraints on the $\Lambda$ scale (in  TeV units) parametrizing the coefficients of the operators as 
$C = 1/\Lambda^2$. Separate bounds for the real and imaginary part of the coefficients are given. When the
bounds are highly asymmetric, separate ones are listed for a positive and a negative value of the coefficient.
}\label{expbounds}
\end{table}

Let us first look at the implications in the  down sector, whose flavor constraints are the strongest.
These are only  coming from the first operator of \eq{4top} that, after rotating to
 the physical basis  using  \eq{vckm},\footnote{In an abuse of notation we will be using the same notation for the quarks in the physical and interaction basis.}
  gives a contribution to the operators
  ${\mathcal Q}_{1}^{sd}$, ${\mathcal Q}_{1}^{bd}$ and ${\mathcal Q}_{1}^{bs}$, as defined in Table~\ref{expbounds},
 with a coefficient 
\be
C({\mathcal Q}_{1}^{sd})\simeq \frac{Y_t^2x_t^2}{\LIR^2}\left[(V_{\textrm{CKM}}^\dagger)_{23}(V_{\textrm{CKM}})_{31}
 \right]^2\simeq 
 10^{-7}\frac{x_t^2}{\LIR^2} e^{i \theta_{\rm CKM}}\, ,
 \label{coeff}
\ee
where $\theta_{\rm CKM}$ denotes the complex phase appearing in the product of the CKM elements, and 
\begin{equation}
 \frac{C({\mathcal Q}_{1}^{bd})}{[(V^\dagger_{\textrm{CKM}})_{33} (V_{\textrm{CKM}})_{31}]^2} 
= \frac{ C({\mathcal Q}_{1}^{bs})}{[(V^\dagger_{\textrm{CKM}})_{33} (V_{\textrm{CKM}})_{32}]^2}
=\frac{C({\mathcal Q}_{1}^{sd})}{[(V_{\textrm{CKM}}^\dagger)_{23}(V_{\textrm{CKM}})_{31} ]^2} \,.
\label{correlation}
\end{equation}
\eq{correlation} leads to interesting consequences.
It predicts    no new phases in $K-\bar K$ and $B-\bar B$  mixing beyond the SM one.
Furthermore,  it  implies that the contributions to the three observables $\epsilon_K$, 
$\Delta M_{B_d}$ and $\Delta M_{B_s}$ are all of the order of the present experimental sensitivity.
Indeed, by looking at the constraints on $\Delta F = 2$ operators reported in Table~\ref{expbounds}, we find
that the three observables $\epsilon_K$,  $\Delta M_{B_d}$ and $\Delta M_{B_s}$ 
give roughly  the same bound. 
The correlation \eq{correlation}  also  arises  in MFV scenarios,  and a bound
has been derived  on the size  of these effects (see Table~\ref{expbounds})
that leads in our case to
\begin{equation}
\LIR \gtrsim 5 x_t\, \mathrm{TeV}\ .\ \ \ \ 
\label{bound1}
\end{equation}
 For $x_t\sim 1/2$ we can accommodate   \eq{bound1} for  values of $\LIR$ 
as low as those needed to pass  EWPT, $\LIR \gtrsim 3$ TeV \cite{Grojean:2013qca,Panico:2015jxa}.
The correlations in~\eq{correlation}  are an interesting smoking gun for these scenarios of flavor, that could be tested in the future with a better determination of the observables.
In particular, we must observe a different value of  $\Delta M_{B_{d,s}}$ from the SM one,
with the ratio fixed:
\be
\frac{\Delta M_{B_d}}{\Delta M_{B_s}}\simeq \left.\frac{\Delta M_{B_d}}{\Delta M_{B_s}}\right|_{\rm SM}\,.
\label{cleancorrelation}
\ee

The impact in the up sector is negligible, since the mixing angles ($\propto Y_{u,c}/Y_t$) are much smaller than in the
down sector. The largest effect comes from the third operator in \eq{4top}, which gives a contribution
\be
C({\mathcal Q}_4^{cu})\simeq \frac{Y_t^2}{\LIR^2}(V^{\rm up}_{R})^*_{32} (V^{\rm up}_{L})_{31}
(V^{\rm up}_{L})^*_{32} (V^{\rm up}_{R})_{31}\sim
\frac{Y_u^2Y_c^2/Y^2_t}{\LIR^2}\simeq
10^{-15}\, \frac{1}{\LIR^2}\, ,
\label{charming}
\ee
where we have taken $\alpha_{L,R}\sim 1$.
This is  many orders of magnitude below  the experimental bound for  $\LIR\sim $ TeV.

Let us now move to the effects at the scales $\Lambda_{f}\gg \LIR$. 
It is clear that contributions at $\Lambda_{b}$ are smaller than those of 
\eq{4top}, as they are suppressed by a larger scale $\Lambda_{b}\gg \LIR$. 
Contributions from  $\Lambda_{c}$ and $\Lambda_{s}$ can however be sizable as they involve second family  quarks.
The most relevant contributions are
\footnote{Notice that contributions to the ${\cal Q}_2$ and $\widetilde {\cal Q}_2$ operators
require two Higgs insertions and are thus highly suppressed.}
\be
\frac{\gIR^2\epsilon^{(2)\, 4}_{c_L}}{\Lambda_c^2}(\overline Q_{L2} \gamma^\mu  Q_{L2})^2\ ,\ \ \  
\frac{\gIR^2\epsilon^{(2)\, 3}_{c_L} \epsilon^{(2)}_{t_L}}{\Lambda_c^2}(\overline Q_{L2} \gamma^\mu  Q_{L3})(\overline Q_{L2} \gamma_\mu  Q_{L2})\ ,\ \ \  
\frac{\gIR^2(\epsilon^{(2)}_{s_L} \epsilon^{(2)}_{s_R})^2}{\Lambda_s^2}(\overline Q_{L2}  s_R)(\overline s_R Q_{L2})\,.
\label{4strange0}
\ee
Using \eq{yukawas}, we can trade the scales $\Lambda_{c,s}$ by $\LIR$,
  and write \eq{4strange0}  for  $d_H=2$ as
\be
\frac{Y_c^2 x_c^2}{\LIR^2}(\overline Q_{L2} \gamma^\mu Q_{L2})^2\ ,\ \ \ 
\frac{Y_c^2 x_c^2 \alpha_L^{ct}}{\LIR^2} (\overline Q_{L2} \gamma^\mu Q_{L3})(\overline Q_{L2} \gamma_\mu  Q_{L2})\ ,\ \ \ 
\frac{Y_s^2}{\LIR^2}(\overline Q_{L2}  s_R)(\overline s_R Q_{L2}   )\,, 
\label{4strange}
\ee
where $x_c=\epsilon^{(2)}_{c_L}/\epsilon^{(2)}_{c_R}$.
After rotating to  the physical basis,  the  operators in \eq{4strange} give  respectively
\bea
C({\mathcal Q}_{1}^{sd})&\simeq& \frac{Y_c^2x_c^2}{\LIR^2}\left[ (V_{\textrm{CKM}}^\dagger)_{22}(V_{\textrm{CKM}})_{21}\right]^2\simeq 
4\times 10^{-7}\,\frac{x_c^2}{\LIR^2}\ ,\label{charmingit}\\
C({\mathcal Q}_{1}^{sd}) &\simeq& \frac{Y_c^2x_c^2 \alpha_L^{ct}}{\LIR^2} (V_{\textrm{CKM}}^\dagger)^2_{22}(V_{\textrm{CKM}})_{21}(V_{\textrm{CKM}})_{31}\simeq 
1.6 \times 10^{-8}\,\frac{x_c^2 \alpha_L^{ct}}{\LIR^2}\ ,\label{charmingit2}\\
C({\mathcal  Q}_{4}^{sd})&\simeq& \frac{Y_s^2}{\LIR^2}
(V^\dagger_{\textrm{CKM}})_{22} (V^{\rm down}_{R})_{21}
(V^{\rm down}_{R})_{22}^* (V_{\textrm{CKM}})_{21}\simeq 
9\times 10^{-10}\, \frac{\alpha_L^{ds}}{\LIR^2}\,.
\label{ats}
\eea
The first contribution is real and therefore only affects $\Delta M_K$, while the other two
can be complex and contribute to  $\epsilon_K$.
Their experimental bounds lead to
\footnote{In computing the bounds on operators generated at $\Lambda_f \gg  \LIR$,
 running effects should also be taken into account. These include
the running of $Y_f$ (which decrease at high energy),
as well as the running of the $4$-fermion effective interactions (which determine a mild increase in the bounds
for the ${\cal Q}_4$ operators). These two effects partially compensate each other. Since the numerical impact
is not large, for simplicity we will not take into account the running in our estimates.
}
\be
\LIR \gtrsim  0.6 x_c\ \mathrm{TeV}\ ,\ \ \ \
\LIR \gtrsim 1.8 x_c \sqrt{\alpha_L^{ct}}\ \mathrm{TeV} \ ,\ \ \ \
\LIR \gtrsim 5\sqrt{\alpha_L^{ds}}\ \mathrm{TeV}\,.
\label{bound2}
\ee
To derive these bounds we have assumed  that the contributions~\eq{charmingit2} and \eq{ats}
 have maximal complex phase $\sim\pi/4$,
as we will assume throughout  the article. 
The bounds in \eq{bound2} are roughly comparable to
the one in \eq{bound1}, and  can be  accommodated  for  $\LIR$ of $few$ TeV.
These extra contributions  to $\epsilon_K$  spoil the correlation in \eq{correlation}, but  preserve \eq{cleancorrelation}. 
Indeed, it is easy to realize that contributions at $\Lambda_{c,s}$ 
to $B$ physics (and also $D$ physics) are negligible.

Finally, we also have contributions arising at $\Lambda_d$.
The most  relevant ones are  those to the operator  
${\mathcal Q}_{4}^{sd}$.
For $d_H=2$ we have
\be
C( {\mathcal Q}_{4}^{sd})\simeq \frac{Y_d^2\alpha^{ds}_L\alpha^{ds}_R}{\LIR^2}\simeq 
9\times 10^{-10}\,\frac{\alpha^{ds}_L}{\LIR^2}\ ,\\
\label{fromld}
\ee
where we have used \eq{alphar}. This contributions are  as sizable as 
\eq{ats}.

  \begin{table}
\footnotesize
\centering
\def\arraystretch{1.25}
\begin{tabular}{@{}c|c|c|c|c|c@{}}
$\Delta F =2$ & $t$ partly-comp. & $s$ partly-comp. &  bilin. mixing (2nd fam.) &  bilin. mixing (1st fam.)  & Anarchic\\
\hline
\hline
\rule{0pt}{1.2em}${\cal Q}_1^{sd}$ &  \boldmath $\LIR \gtrsim 5 x_t$ & \boldmath $\LIR \gtrsim 4 x_t$ &  $\LIR \gtrsim 1.8 x_c \sqrt{\alpha_L^{ct}}$ & $\LIR \gtrsim 0.2 x_d$ & \boldmath $\LIR \gtrsim 4 x_t$\\
\rule{0pt}{1.2em}${\cal Q}_2^{sd}$ & -- & $\LIR \gtrsim 1 \sqrt{g_*}$ & $\cdot$ & $\cdot$ & $\LIR \gtrsim 1 \sqrt{g_*}$\\
\rule{0pt}{1.4em}$\widetilde {\cal Q}_2^{sd}$ & -- & $\LIR \gtrsim 0.5 \sqrt{g_* \alpha_L^{ds}}$ &  $\cdot$ & $\cdot$ & $\LIR \gtrsim 1 \sqrt{g_*}$\\
\rule[-.5em]{0pt}{1.9em}${\cal Q}_4^{sd}$ & -- & \boldmath $\LIR \gtrsim 5  \sqrt{\alpha_L^{ds}}$ & \boldmath $\LIR \gtrsim 5 \sqrt{\alpha_L^{ds}}$ & \boldmath $\LIR \gtrsim 5 \sqrt{\alpha_L^{ds}}$ & \boldmath $\LIR \gtrsim 10 $\\
\hline
\rule{0pt}{1.2em}${\cal Q}_1^{bd}$ & \boldmath $\LIR \gtrsim 5 x_t$ & \boldmath $\LIR \gtrsim 6 x_t$ &  $\cdot$ & $\cdot$ & \boldmath $\LIR \gtrsim 6 x_t$\\
\rule{0pt}{1.2em}$\widetilde {\cal Q}_2^{bd}$ & -- & $\LIR \gtrsim 0.3 \sqrt{g_* \alpha_L^{ds}}$ &  $\cdot$ & $\cdot$ & $\LIR \gtrsim 0.6 \sqrt{g_*}$\\
\rule[-.5em]{0pt}{1.9em}${\cal Q}_4^{bd}$ & -- & $\LIR \gtrsim 0.4 \sqrt{\alpha_L^{sd}}$ & $\LIR \gtrsim 0.3 \sqrt{\alpha_L^{db}}$ &  $\cdot$ & $\LIR \gtrsim 0.8 $\\
\hline
\rule{0pt}{1.2em}${\cal Q}_1^{bs}$ & \boldmath $\LIR \gtrsim 5 x_t$ & \boldmath $\LIR \gtrsim 7 x_t$ &  $\LIR \gtrsim 0.6 \alpha_R^{cb} x_c$ & $\cdot$ & \boldmath $\LIR \gtrsim 7 x_t$\\
\rule{0pt}{1.4em}$\widetilde {\cal Q}_2^{bs}$ & -- & $\LIR \gtrsim 0.4 \sqrt{g_*}$  & $\cdot$ & $\cdot$ & $\LIR \gtrsim 0.4 \sqrt{g_*}$\\
\rule[-.5em]{0pt}{1.9em}${\cal Q}_4^{bs}$ & -- & $\LIR \gtrsim 1 $ &  $\LIR \gtrsim 0.1 \sqrt{\alpha_L^{sb} }$ & $\cdot$ & $\LIR \gtrsim 1 $\\
\hline
\rule{0pt}{1.2em}${\cal Q}_1^{cu}$ & $\cdot$ & $\cdot$ &  $\cdot$ & $\cdot$ & $\LIR \gtrsim 1 x_t$\\
\rule{0pt}{1.2em}${\cal Q}_2^{cu}$ & $\cdot$ & $\cdot$ &  $\cdot$ & $\cdot$ & $\LIR \gtrsim 0.7 \sqrt{g_*}$\\
\rule[-.5em]{0pt}{1.7em}${\cal Q}_4^{cu}$ & $\cdot$ & $\cdot$ &  $\cdot$ & $\cdot$ & $\LIR \gtrsim 1.1$\\
\end{tabular}
\caption{\it Bounds on $\LIR$ for the different scenarios considered in the text.
The effects are separated according to their origin: from the top (or strange) partial compositeness at $\LIR$,
or from  the UV scale $\Lambda_f$ at which the second and first families get bilinear mixings  to the Higgs.
 The results are given in  TeV.
  Entries with a "$\cdot$" correspond to negligible bounds, while "--" means that the corresponding
operator is not generated. The most relevant constraints are highlighted in boldface.}\label{tab:bounds1}
\end{table}

The above conclusions however drastically depend  on  $d_H$.
For $d_H>2$ we have that   the contributions from $\Lambda_{c,s,d}$
are enhanced,  with  respect  to  Eqs.~(\ref{charmingit})--(\ref{ats}) and \eq{fromld},
by a factor
$\left({\Lambda_{f}}/{\LIR}\right)^{2d_H-4}$. 
Therefore $d_H>2$ can only be consistent with the experimental bounds if
$\Lambda_{f}\sim\LIR$ that  
corresponds to  the anarchic scenario. 
This implies that generating the mass for the charm, strange or down
from bilinear mixing at $\Lambda_{f}\gg \LIR$ is only possible for $d_H\lesssim 2$.

\subsection{Neutron EDM}

Dipole operators can also be induced at  $\Lambda_f$.
These operators are strongly constrained, in particular from the measurement of the neutron EDM, which
place a bound on quark dipole operators of the form
\begin{equation}
c_{\rm edm}^q\, \bar q_L \sigma^{\mu\nu} g_s G_{\mu\nu} q_R\,,
\label{qedm}
\end{equation}
or analogous operators involving the photon field-strength (see Table~\ref{expbounds}). 
In the anarchic case the current measurements lead to  very severe bounds,
 $\LIR \gtrsim 48\, (\gIR/4\pi)$ TeV from the down-quark EDM, and $\LIR  \gtrsim 18\, (\gIR/4\pi)$ TeV  from  the up-quark EDM.
These bounds were calculated under the assumption  that dipole operators are induced   at the one-loop level 
and therefore must
carry a factor $\gIR^2/16\pi^2$ \cite{KerenZur:2012fr}, as it occurs in holographic descriptions of the model \cite{Agashe:2004cp}.
Obviously, for maximal coupling  $\gIR\sim 4\pi$ this loop factor  is  of  order one, 
not  introducing  any extra suppression. Hereafter we will also follow this assumption for our estimates.

In our scenarios for flavor the contributions to $c_{\rm edm}^{u,d}$ 
are all very small, due to  either small mixings or a  large scale $\Lambda_f$ suppressing the processes.
In fact, the main contribution to the neutron EDM comes from a top EDM that  can be induced at $\LIR$
with a size
\be
c_{\rm edm}^t\simeq \frac{\gIR^2}{16\pi^2}\frac{m_t}{\LIR^2}\,.
\label{topcon}
\ee
According to the bound in Table~\ref{expbounds}, we obtain $\LIR\gtrsim 3 (\gIR/4\pi)$~TeV,
implying that we expect in these scenarios 
a neutron EDM  below, but not much smaller than,    its present experimental limit.

Contributions originating  at  $\Lambda_f$ are much smaller.
The reason is that  EDM operators must involve the Higgs field that at high energies
corresponds to the composite operator ${\cal O}_H$ of dimension larger than one. Therefore 
the contribution to EDMs is suppressed by  $d_H+1$ powers of $\Lambda_f$. For example,  at  $\Lambda_b$,  
we expect a bottom-EDM from the operator
\be
\frac{\gIR^2}{16\pi^2}\frac{\epsilon^{(3)}_{b_L}\epsilon^{(3)}_{b_R}}{\Lambda_b^{d_H+1}}\, 
\bar Q_{L3} {\cal O}_H   \sigma^{\mu\nu} g_s G_{\mu\nu} b_R\,,
\label{opedm}
\ee
which gives
\be
c_{\rm edm}^b\simeq \frac{\gIR^2}{16\pi^2}
\frac{m_b}{\Lambda_b^2}
\,.
\ee
This   is  much smaller than present bounds unless $\Lambda_b\sim \LIR$.

\subsection{\boldmath $\Delta F=1$  transitions}

Similarly to EDMs, contributions to  flavor dipole transitions can also be present, the most relevant ones being
$\overline s_{R,L} \sigma^{\mu\nu} e F_{\mu\nu} b_{L,R}$  that contributes to  $b\to s\gamma$, 
and 
$\overline s_{R,L} \sigma^{\mu\nu} g_s G_{\mu\nu} d_{L,R}$
that contributes to $\epsilon'/\epsilon$. The estimates of these effects are similar to the ones for the neutron EDM in \eq{opedm},
leading to  small contributions to these observables.

There are  also non-dipole contributions to $\Delta F=1$  transitions arising from operators like 
$\bar s_L\gamma^\mu d_L H^\dagger \overleftrightarrow  D_\mu H$
that on the EWSB vacuum give flavor-changing $Z$-couplings,
which are severely constrained by $K_L\to \mu^+\mu^-$ and $\epsilon'/\epsilon$,
or equivalent operators with the bottom, $\bar s_L\gamma_\mu b_L H^\dagger \overleftrightarrow D_\mu H$, which give
contributions to the processes $B\to \ell^+\ell^-,X \ell^+\ell^-$. 
The largest  contribution  arises  from  top operators induced at $\LIR$ that give
\bea
&&\frac{ (\gIR\epsilon^{(3)}_{t_L})^2}{\LIR^2}\bar Q_{L3}\gamma^\mu Q_{L3} i H^\dagger \overleftrightarrow D_\mu H
\simeq
\frac{\gIR Y_tx_t}{\LIR^2}
\Big((V_{\rm CKM}^\dagger)_{33}\, \bar b_L  + (V_{\rm CKM}^\dagger)_{23}\, \bar s_L +(V_{\rm CKM}^\dagger)_{13}\, \bar d_L \Big)\gamma^\mu \nonumber\\
&&\times
\Big((V_{\rm CKM})_{33}\, b_L+(V_{\rm CKM})_{32}\, s_L+(V_{\rm CKM})_{31}\,  d_L\Big) i H^\dagger \overleftrightarrow D_\mu H\,,
\label{df1}
\eea
similarly to the anarchic case.
Interestingly, \eq{df1} shows that the contributions to $K_L\to \mu^+\mu^-$ (and $\epsilon'/\epsilon$), $B\to (X)\ell\ell$
and corrections to $Z\bar b_Lb_L$ are  correlated  and all are close to the 
experimental bounds; we obtain respectively the constraints
\be
\LIR \gtrsim 4\, \sqrt{\gIR x_t}\ \mathrm{TeV}\ ,\ \ \ \ 
\LIR \gtrsim 3\, \sqrt{\gIR x_t}\ \mathrm{TeV}\ ,\ \ \ \ 
\LIR \gtrsim 5\, \sqrt{\gIR x_t}\ \mathrm{TeV}\,.
\ee
We must point out however that there is another  dimension-six operator contributing to these observables, 
$\bar Q_{L3}\sigma^a\gamma^\mu Q_{L3} H^\dagger \sigma^a \overleftrightarrow D_\mu H$,
that in the case of a custodial $P_{LR}$  symmetry in the strong sector
  cancels the contribution from \eq{df1}~\cite{Agashe:2006at}. 
 This symmetry is present in simple models of composite Higgs
and for this  reason these effects could be further suppressed.

Finally, there can be also contributions to operators like
$\bar s_L\gamma_\mu d_L D_\nu F_Z^{\mu\nu}$,  where $F_Z^{\mu\nu}$ is the field-strength 
of the $Z$. These operators, however, are suppressed by a factor $g^2/\gIR^2$ with respect to those in \eq{df1}.

\begin{table}
\footnotesize
\centering
\def\arraystretch{1.25}
\begin{tabular}{@{\;}c|c|c|c|c@{\;}}
\rule{0pt}{1.1em}$\Delta F =1$  & $t$ partly comp. & $b$ partly comp. & $s$ partly comp. & Anarchic\\
\hline
\hline
\rule{0pt}{1.45em}$\overline s_L \sigma^{\mu\nu} e F_{\mu\nu} b_R$ & -- & $\LIR \gtrsim 0.12 \gIR$ &  $\LIR \gtrsim 0.12 \gIR$ & $\LIR \gtrsim 0.12 \gIR$\\
\rule[-.5em]{0pt}{1.7em}$\overline s_R \sigma^{\mu\nu} e F_{\mu\nu} b_L$ & -- & $\cdot$ & \boldmath $\LIR \gtrsim 0.8 \gIR$ & \boldmath $\LIR \gtrsim 0.8 \gIR$\\
\hline
\rule{0pt}{1.45em}$\overline s_L \sigma^{\mu\nu} g_s G_{\mu\nu} d_R$ & -- & $\cdot$ & $\LIR \gtrsim 0.5 \gIR$ & \boldmath $\LIR \gtrsim 1.1 \gIR$\\
\rule[-.5em]{0pt}{1.7em}$\overline s_R \sigma^{\mu\nu} g_s G_{\mu\nu} d_L$ & -- & $\cdot$ & \boldmath $\LIR \gtrsim 1.1 \gIR$ & \boldmath $\LIR \gtrsim 1.1 \gIR$\\
\hline
\rule[-.5em]{0pt}{2.em}$\overline s_L \gamma^\mu b_L H^\dagger i \overleftrightarrow{D}_\mu H$ & \boldmath $\LIR \gtrsim 3 \sqrt{g_* x_t}$ (*) & $\LIR \gtrsim 0.4 \sqrt{g_* x_b}$ & $\LIR \gtrsim 0.4 \sqrt{g_* x_b}$ & \boldmath $\LIR \gtrsim 3 \sqrt{g_* x_t}$\\
\rule[-.75em]{0pt}{2.4em}$\overline s_L \gamma^\mu d_L H^\dagger i \overleftrightarrow{D}_\mu H$ & \boldmath $\LIR \gtrsim 4 \sqrt{g_* x_t}$ (*) & $\LIR \gtrsim 0.50 \sqrt{g_* x_b}$ & $\LIR \gtrsim 0.5 \sqrt{g_* x_b}$ & \boldmath $\LIR \gtrsim 4 \sqrt{g_* x_t}$\\
\hline
\hline
\rule{0pt}{1.1em}$\Delta F =0$ & $t$ partly-comp. & $b$ partly-comp. & $s$ partly-comp. & Anarchic\\
\hline
\hline
\rule[-.75em]{0pt}{2.2em}$\overline b_L \gamma^\mu b_L H^\dagger i \overleftrightarrow{D}_\mu H$ & \boldmath $\LIR \gtrsim 5 \sqrt{g_* x_t}$ (*)  & $\LIR \gtrsim 0.6 \sqrt{g_* x_b}$ & $\LIR \gtrsim 0.6 \sqrt{g_* x_b}$ & \boldmath $\LIR \gtrsim 5 \sqrt{g_* x_t}$\\
\hline
\hline
\rule{0pt}{1.1em}Neutron EDM & $t$ partly-comp. & $b$ partly-comp. & $s$ partly-comp. & Anarchic\\
\hline
\hline
\rule{0pt}{1.75em}$\overline d_L \sigma^{\mu\nu} e F_{\mu\nu} d_{R}$ & -- & $\LIR \gtrsim 0.24 \gIR \sqrt{\alpha_L^{db}}$ & \boldmath $\LIR \gtrsim 1.2 \gIR \sqrt{\alpha_L^{ds}}$ & \boldmath $\LIR \gtrsim 2.5 \gIR$\\
\rule{0pt}{1.6em}$\overline u_L \sigma^{\mu\nu} e F_{\mu\nu} u_{R}$ & $\cdot$ & $\cdot$ & $\cdot$ & \boldmath $\LIR \gtrsim 0.9 \gIR$\\
\rule{0pt}{1.6em}$\overline d_L \sigma^{\mu\nu} g_s G_{\mu\nu} d_{R}$ & -- & $\LIR \gtrsim 0.3 \gIR \sqrt{\alpha_L^{db}}$ & \boldmath $\LIR \gtrsim 1.5 \gIR \sqrt{\alpha_L^{ds}}$ & \boldmath $\LIR \gtrsim 3.2 \gIR$\\
\rule{0pt}{1.6em}$\overline u_L \sigma^{\mu\nu} e F_{\mu\nu} u_{R}$ & $\cdot$ & $\cdot$ & $\cdot$ & \boldmath $\LIR \gtrsim 1.2 \gIR$\\
\rule{0pt}{1.6em}$\overline c_L \sigma^{\mu\nu} g_s G_{\mu\nu} c_{R}$ & $\cdot$ & $\cdot$ & $\cdot$ & \boldmath $\LIR \gtrsim 1 \gIR$\\
\rule{0pt}{1.6em}$\overline b_L \sigma^{\mu\nu} e F_{\mu\nu} b_{R}$ & -- & $\LIR \gtrsim 0.6 \gIR$ & $\cdot$ & $\LIR \gtrsim 0.6 \gIR$\\
\rule[-.75em]{0pt}{2.35em}$\overline t_L \sigma^{\mu\nu} e F_{\mu\nu} t_{R}$ & $\LIR \gtrsim 0.24 \gIR$ & $\cdot$ & $\cdot$ & $\LIR \gtrsim 0.24 \gIR$\\
\hline
\hline
\rule[-.5em]{0pt}{1.6em}Leptons & $t$ party comp. & $\tau$ partly-comp. & $\mu$ partly-comp. & Anarchic\\
\hline
\hline
\rule{0pt}{1.4em}$\overline e_L \sigma^{\mu\nu} e F_{\mu\nu} e_R$ & \boldmath $\LIR \gtrsim 1.6 \sqrt{g_* x_t}$ & $\LIR \gtrsim 0.5 \gIR \sqrt{\alpha_{L}^{e \tau}\alpha_{R}^{e \tau}}$ & \boldmath $\LIR \gtrsim 2 \gIR \sqrt{\alpha_{L}^{e \mu}\alpha_{R}^{e \mu}}$ & \boldmath $\LIR \gtrsim 32 \gIR$\\
\rule{0pt}{1.4em}$\overline \mu \sigma^{\mu\nu} e F_{\mu\nu} e_{L,R}$ & $\cdot$ & \boldmath $\LIR \gtrsim 1.2 \gIR \sqrt{\alpha_{L,R}^{e \tau}\alpha_{R,L}^{\mu \tau}}$ & \boldmath $\LIR \gtrsim 5 \gIR \sqrt{\alpha_{L,R}^{e \mu}}$ & \boldmath $\LIR \gtrsim 19 \gIR$\\
\rule{0pt}{1.4em}$\overline \tau \sigma^{\mu\nu} e F_{\mu\nu} \mu_{L,R}$ & $\cdot$ & $\LIR \gtrsim 0.7 \gIR \sqrt{\alpha_{L,R}^{\mu \tau}}$ &  \boldmath $\LIR \gtrsim 1.3 \gIR$ & \boldmath $\LIR \gtrsim 1.3 \gIR$\\
\rule{0pt}{1.4em}$\overline \tau \sigma^{\mu\nu} e F_{\mu\nu} e_{L,R}$ & $\cdot$ & $\cdot$ & $\LIR \gtrsim 0.1 \gIR \sqrt{\alpha_{L,R}^{e \mu}}$ & $\LIR \gtrsim 0.4 \gIR$
\end{tabular}
\caption{\it Bounds on $\LIR$ from assuming that  the top, bottom, etc.~are partly composite at $\LIR$.
The results are given in TeV. 
Entries with a "$\cdot$" correspond to negligible bounds, while "--" means that the corresponding
operator is not generated. The most relevant constraints are highlighted in boldface.
If a custodial $P_{LR}$ symmetry \cite{Agashe:2006at} is present in the top mixings, the bounds denoted by (*) are absent.}\label{tab:bounds2}
\end{table}

\subsection{Electron EDM, \boldmath $\mu\to e\gamma$ and $\tau\to \mu\gamma$}

Assuming that the  origin of the lepton  masses is   the same  one as for the down-type quark masses 
described above, 
we expect  ${\cal Y}_{\rm lepton}$  and the rotation matrices to have the same structure as
 \eq{matrixYd}  and \eq{rotationLR} respectively,
 with the obvious replacement $d,s,b\to e,\mu,\tau$. 
The corresponding $\alpha_{L,R}$  for the lepton sector 
are free parameters, that we will take to be order one for our estimates.

The main experimental constraints on   possible effective operators induced at the scales $\Lambda_{e,\mu,\tau}$
are  the electron EDM,  $\mu\to e\gamma$  and $\tau\to \mu\gamma$, that come  from similar dipole structures:
\be\label{eq:lepton_dipole}
c_{\rm edm}^e\, \overline e_L \sigma^{\mu\nu} e F_{\mu\nu} e_R\ , \ \ \ \
c_{\rm meg}\, \overline e_L \sigma^{\mu\nu} e F_{\mu\nu} \mu_R\ , \ \ \ \
c_{\rm tmg}\, \overline \mu_L \sigma^{\mu\nu} e F_{\mu\nu} \tau_R
\,,
\ee
and analogous ones obtained interchanging the chiralities, $L\leftrightarrow R$.
In the anarchic case  the first  two  operators in \eq{eq:lepton_dipole} put the most severe constraints  (see Table~\ref{tab:bounds2}).
In our scenarios, 
however, we find that these contributions are  very small for the same reason as for the neutron EDM.
The largest contribution  arises at $\Lambda_\tau$, and give for $d_H=2$
\be
c_{\rm edm}^e\simeq  \left(\frac{\gIR}{4\pi}\right)^2(V_L^{\rm lepton})^*_{31}(V_R^{\rm lepton})_{31}\, 
\frac{\gIR v\LIR}{\Lambda_\tau^3}\sim
\left(\frac{\gIR}{4\pi}\right)^2\frac{Y_eY_\tau}{\gIR^2}\, 
\frac{m_e}{\LIR^2}\,,
\ee
which is extremely small.
Similarly, for  $\mu\to e\gamma$ and $\tau\to \mu\gamma$, we get  at $\Lambda_\tau$: 
\bea
c_{\rm meg}&\simeq&
\left(\frac{\gIR}{4\pi}\right)^2(V_L^{\rm lepton})^*_{32}(V_R^{\rm lepton})_{31}\, 
\frac{\gIR v\LIR}{\Lambda_\tau^3}
\sim
\left(\frac{\gIR}{4\pi}\right)^2  \frac{Y_eY_\tau}{\gIR^2}\frac{m_\mu}{\LIR^2}\,,\\
c_{\rm tmg}&\simeq&
\left(\frac{\gIR}{4\pi}\right)^2(V_L^{\rm lepton})^*_{32}\, 
\frac{\gIR v\LIR}{\Lambda_\tau^3}
\sim
\left(\frac{\gIR}{4\pi}\right)^2  \frac{Y_\mu Y_\tau}{\gIR^2}\frac{m_\tau}{\LIR^2}\,,
\eea
that are several orders of magnitude below the experimental bound.

\begin{figure}[t]
\centering
\includegraphics[width=0.375\textwidth]{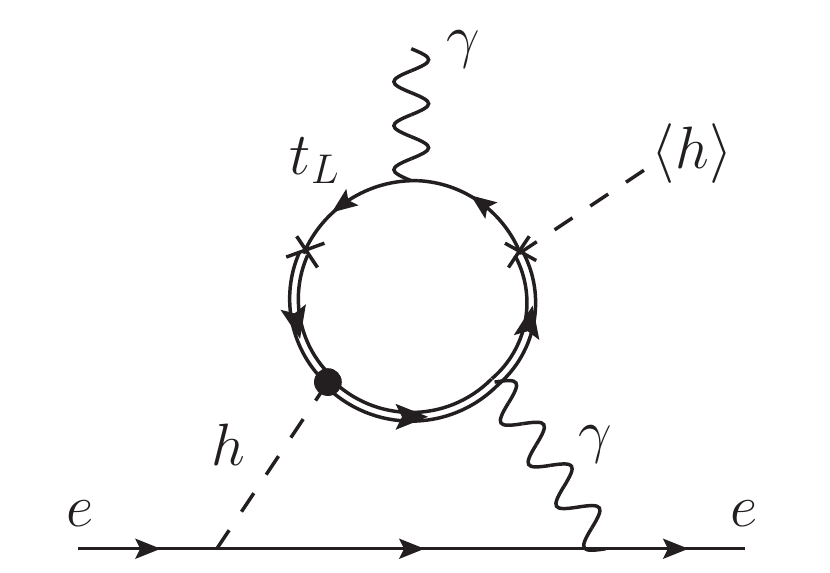}
\caption{\it 
A representative two-loop contribution to the electron EDM.
The   double-line represents a resonance from the strong sector.
}
\label{fig:bz}
\end{figure}

Additional contributions to the electron EDM can come from Barr--Zee-type $2$-loop diagrams     \cite{Barr:1990vd}
as shown in Fig.~\ref{fig:bz}.
These involve   CP-violating one-loop induced vertices such as $H^\dagger D^2_\rho H  \tilde F_{\mu\nu}F^{\mu\nu}$
arising from the strong sector, mainly from a loop of top resonances.\footnote{There is also the possibility to have
a vertex  involving a $Z$,  but this contribution to the EDM is suppressed
 as a consequence of $C$ invariance that makes only the (very small) vector part of the $Z$ coupling to the electron to contribute 
 \cite{Barr:1990vd}.}
The estimate of the size of these couplings are very model dependent.
In the particular motivated case of a pseudo-Nambu--Goldstone boson (PNGB)  Higgs
these couplings cannot be generated from the strong sector alone, as they are protected by the global symmetry under which the
Goldstone Higgs transforms. 
Therefore we need a SM particle to be involved in the loop.
We can take as an estimate the contribution involving the $t_L$ (see  Fig.~\ref{fig:bz})
that induces the vertex $H^\dagger D^2_\rho H  \tilde F_{\mu\nu}F^{\mu\nu}$
with a coefficient  $\sim e^2x_tY_t\gIR/(16\pi^2)$ (omitting powers of $\LIR$).
Using the results of Ref.~\cite{Brod:2013cka}, in which the
Barr--Zee contribution to the electron EDM is computed in the presence of CP-violating Higgs interactions to the top,
$-i \widetilde \kappa_t Y_t (\overline t \gamma_5 t) h/\sqrt{2}$,
and found 
$|\widetilde \kappa_t| < 0.01$,
we have, after the proper rescaling for our case,
\begin{equation}
 \frac{x_tY_t \gIR}{\LIR^2}
\lesssim 
0.01\frac{Y^2_t}{m^2_t}
\,,
\end{equation}
 that leads to  the  bound
\begin{equation}
\LIR \gtrsim 1.6\, \sqrt{{\gIR x_t}}\ \textrm{TeV}\,.
\end{equation}
The size of this correction is thus comparable with the present experimental bounds and should be visible in future experiments.
Notice that in the cases in which the Higgs is not a PNGB, this effect is enhanced by a factor $\gIR/Y_t$.

Barr--Zee-type contributions to $\mu\to e\gamma$
are also sizeable in anarchic  models   \cite{Beneke:2015lba}, but in our scenarios for flavor these contributions are  
very small since the Higgs  flavor-changing  couplings to  leptons are highly suppressed
--see Section~\ref{secHff}.

\section{Alternative scenarios}

Although so far we considered a scenario in which  the different fermion masses arise at different 
UV scales $\Lambda_f$, we could also consider simpler cases with fewer UV scales or with more particles than the top
with masses arising from partial compositeness at $\LIR$.
In the following we present several alternative scenarios pointing out in which cases 
there is a clash with the experimental bounds.

\begin{itemize}
\item {\bf  First-family  masses generated at the same  scale $\Lambda_1$:}
  
  We could take the economical  assumption that all first-family fermion masses arise at the same scale  $\Lambda_1\sim \Lambda_d\sim 3\times 10^{8}$ GeV, corresponding to the scale of the heaviest fermion, the down quark.
 The fact that $m_e < m_u < m_d$ could be accommodated in this case by taking the mixing terms 
 $\epsilon^{(1)}_{e_R,e_L}$ and  $\epsilon^{(1)}_{u_R}$ to be slightly smaller  than one. 
  None of the  estimates made in the previous section  are changed in this case.
The reason is that none of the main contributions were originating at $\Lambda_{u}$ or $\Lambda_e$,
as these were very small.

\item {\bf  Second-family  masses generated at the same  scale $\Lambda_2$:}

Similarly, we could  assume that  all second-family fermions get their masses at one single   scale
$\Lambda_2\sim \Lambda_c\sim 10^{6}$ GeV.
Again,  it is easy to show that the  estimates of the previous section are not affected.
Of course,   contributions at the scale $\Lambda_2$ to up quark  and electron EDM, as well as $\mu\to e\gamma$ are larger now  as $\Lambda_2\gg \Lambda_{s,\mu}$, but these are still  few orders of magnitude below the experimental bounds.
Contributions to $\Delta F =2$ $4$-fermion interactions are however not affected, since
for $d_H \simeq 2$ they can be written, using  \eq{yukawas},
as a function of $Y_s$ and  $\LIR$, independently of  $\Lambda_s$.

\item {\bf  Partly-composite third-family fermions at $\LIR$:}

Following the above approach of  family reunion,
 we can consider  the case in which all third family fermions are, analogously to the top, partly composite, i.e.,
having their masses arising at $\LIR$.

\begin{itemize}

\item \underline{Partly-composite bottom}: In this case there are new contributions to $\Delta F=2$ that have
the same structure as \eq{4top} but with the replacement $t_R\to b_R$ and $Y_t\to Y_b$. Due to the $Y_b$ suppression,
one gets contributions much smaller than the present bounds.
There is also now a larger contribution to the bottom-quark EDM, arising at $\LIR$:
\be
c_{\rm edm}^b\simeq \left(\frac{\gIR}{4\pi}\right)^2\frac{m_b}{\LIR^2}\,,
\label{bottomcon}
\ee
which saturates the experimental bound for $\LIR\sim 7\, (\gIR/4\pi)$ TeV.
Additional contributions to the $b \rightarrow s$ and $s \rightarrow d$ transitions as well as to the $Zb\overline b$ coupling
are also generated, which are slightly suppressed with respect to the ones coming from the top partial compositeness
(see Table~\ref{tab:bounds2}).

\item \underline{Partly-composite tau}:  In this case the most relevant observable is
  $\mu\to e\gamma$  that receives  at  $\LIR$ a contribution of order
\bea
c_{\rm meg}&\simeq&
\left(\frac{\gIR}{4\pi}\right)^2\frac{m_\tau}{\LIR^2} (V_L^{\rm lepton})^*_{23}(V_R^{\rm lepton})_{13}\, 
\simeq
\alpha_L^{\mu\tau}\alpha_R^{e\tau}\left(\frac{\gIR}{4\pi}\right)^2  \frac{Y_e}{Y_\tau}\frac{m_\mu}{\LIR^2}\nonumber\\
&\simeq&
3\times 10^{-4}\alpha_L^{\mu\tau}\alpha_R^{e\tau}\left(\frac{\gIR}{4\pi}\right)^2\frac{m_\mu}{\LIR^2}
\,,
\label{megmeg}
\eea
and a similar contribution with $\alpha_R\leftrightarrow \alpha_L$.
From  \eq{megmeg} and the experimental bound in Table~\ref{expbounds}, we get
 \be
 \LIR\gtrsim 15\,\sqrt{\alpha_L^{\mu\tau}\alpha_R^{e\tau}}\left(\frac{\gIR}{4\pi}\right)\ {\rm TeV}\, ,
 \ee
which shows that these corrections can be close to the experimental bound, motivating a better measurement
of $\mu\to e\gamma$ as a probe for this scenario.
Similarly, the electron EDM  and $\tau\to\mu\gamma$ are also enhanced if the tau is partly composite,
leading to the estimates 
 \be
 c^e_{\rm edm}\simeq  \frac{\alpha_L^{e\tau}}{\alpha_L^{\mu\tau}}\frac{m_e}{m_\mu}c_{\rm meg}\ ,\ \ \  \ 
 c_{\rm tmg}\simeq  \frac{1}{\alpha_R^{e\tau}}\frac{Y_\mu}{Y_e}\frac{m_\tau}{m_\mu}c_{\rm meg}\,,
 \label{ratio}
\ee
which saturates the present experimental bounds respectively for
\be
 \LIR\gtrsim 7\,\sqrt{\alpha_L^{e\tau}\alpha_R^{e\tau}}\left(\frac{\gIR}{4\pi}\right)\ {\rm TeV}\ ,\ \ \ \
 \LIR\gtrsim 8\,\sqrt{\alpha_L^{\mu\tau}}\left(\frac{\gIR}{4\pi}\right)\ {\rm TeV}\,.
\ee
Similar bounds apply for $\alpha_R\leftrightarrow \alpha_L$.

\end{itemize}
In summary, if all the third-family fermions are  partly composite at $\LIR$, we could in the near future  see
a positive result from searches for neutron and electron EDM, $\mu\to e\gamma$ or  $\tau\to\mu\gamma$.

\item {\bf  Partly-composite second-family fermions  at $\LIR$:}

As a last example, it can be instructive to consider a case where all except the first-family fermions get their mass
from partial compositeness at $\LIR$.

\begin{itemize}

\item \underline{Partly-composite charm}:  
If the charm is partly composite,  there are   new contributions to $\epsilon_K$, but they go exactly as those in
\eq{charmingit}.  The are also larger contributions to $\Delta M_D$. We find that they can be a factor  
$Y^2_t/Y_c^2\sim 10^5$ larger than those  in \eq{charming}, nevertheless they are  still  below  the experimental bound. 
 The most important new contribution arises for the charm-EDM:
 \be
c_{\rm edm}^c\simeq \left(\frac{\gIR}{4\pi}\right)^2\frac{m_c}{\LIR^2}\,,
\label{bottomcon2}
\ee
which saturates the experimental bound for $\LIR\sim 13\, (\gIR/4\pi)$ TeV.

\item \underline{Partly-composite strange}: 
In this scenario we find the same contribution as in the anarchic case in $K$ physics, shown in Table~\ref{tab:bounds2}. 
Sizable contributions to the down-quark EDM are also generated:
\be
c_{\rm edm}^d\simeq \left(\frac{\gIR}{4\pi}\right)^2\frac{m_s}{\LIR^2}(V_{\rm CKM})^*_{21}(V_R^{\rm down})_{12}
\simeq   0.2\, \alpha_L^{ds}   \left(\frac{\gIR}{4\pi}\right)^2  \frac{m_d}{\LIR^2}\, ,
\ee
which leads to  the  bound $\LIR\gtrsim 19\, (\gIR/4\pi)\sqrt{\alpha_L^{ds}}$~TeV.

\item \underline{Partly-composite muon}: In this case the estimate for the contribution to $\mu\to e\gamma$ and electron EDM
are enhanced with respect  to  those to the partly-composite tau (see  \eq{megmeg} and \eq{ratio})  
by a  factor $Y_\tau/Y_\mu\sim 17$. 
This pushes the bound on $\LIR$ beyond the TeV scale, dominantly due  to $\mu\to e\gamma$.

\end{itemize}
We  conclude  that the option with partly-composite second family  at $\LIR$ seems 
  disfavored by the present experimental data, mainly due to EDMs and $\mu\to e\gamma$. A summary of all bounds is presented in Tables~\ref{tab:bounds1} and \ref{tab:bounds2}.
\end{itemize}

\section{Higgs couplings to fermions}
\label{secHff}

The predictions for  Higgs couplings  depend on the  origin of the fermions masses.
Here we will present the predictions for the models of flavor considered above.
We will focus on  models in which the Higgs arises as a PNGB from the strong sector.
 These  models,  motivated by the lightness of the Higgs,  are able to provide quantitative  predictions
depending only  on how  the global group ${\cal G}$ of the strong sector is broken.
We will consider  in particular the MCHM based on  the   ${\cal G/H}=SO(5)/SO(4)$ coset~\cite{Agashe:2004rs}.
Either in the   case of partly-composite fermions at $\LIR$ or at a larger scale $\Lambda_f$, 
the Higgs couplings  depend  on  how the symmetry $\cal G$ is broken by \eq{linearmix},
and this is determined by specifying how  ${\cal O}_{f_i}$  is embedded into a representation of $\cal G$.
Therefore for  both cases,  the Higgs couplings to fermions  can be written as 
\be
\frac{g^h_{ff}}{g^{h\, \rm SM}_{ff}}=\frac{1-(1+n)v^2/f_h^2}{\sqrt{1-v^2/f_h^2}}\,,
\label{hcoup}
\ee
where $n=0,1,2,...$ and $f_h$ is the Higgs decay constant, $f_h\sim \LIR/\gIR$.
For  ${\cal O}_{f_i}\in {\bf 4}$   (or ${\bf 5}$) of $SO(5)$,    as in the MCHM4 (MCHM5), one finds $n=0$ ($n=1$)~\cite{Giudice:2007fh,Pomarol:2012qf}.
This is also  the case even  if fermion masses come from  bilinears $\bar f_L {\cal O}_H f_R$
with unknown UV origin. Indeed, in this case we need to specify into which representation  of ${\cal G}$
we embed  ${\cal O}_H$,  or, equivalently, to specify an embedding for  $\bar f_L f_R$. This latter 
 can be formally written as a product of the representations of the individual embeddings for $\bar f_L$ and $f_R$.
 Therefore, by specifying these individual embeddings,  we can determine again the Higgs couplings.
As an example, let us consider  ${\cal O}_H\in {\bf 5},{\bf 14}$.
Since $\bf 5\in  \bf \bar 4\times\bf 4$ and $\bf 14\in \bf 5\times\bf 5$, 
we find respectively  $n=0,1$, as in the MCHM4 and MCHM5.

It is also  interesting   to  analyze  the predictions for flavor-changing Higgs couplings.
The coupling $h\bar \tau\mu$ is  of special interest, as this is  the one which experimental constraints have been presented
from $h\to \tau\mu$ \cite{Khachatryan:2015kon,Aad:2015gha}.
We find  however that contributions to this coupling are  very small.  
For example,  even for  the case of a $\tau$  partly-composite at $\LIR$,  we get 
\be
BR(h\to \mu\tau)\simeq \left(\frac{\gIR^2v^2}{\LIR^2}\frac{m_\mu}{m_\tau}\right)^2BR(h\to \tau\tau)\sim 2\times 10^{-4}\left(\frac{\gIR v}{\LIR}\right)^4\, ,
\ee
that is much below the present limit $BR(h\to \mu\tau)<1.51 \%$  from CMS \cite{Khachatryan:2015kon}
($1.85\%$ from ATLAS \cite{Aad:2015gha}).
A  larger effect is found if  $\mu$  is partly composite at $\LIR$:
\be
BR(h\to \mu\tau)\simeq \left(\frac{\gIR^2v^2}{\LIR^2}\sqrt{\frac{m_\mu}{m_\tau}}\,\right)^2BR(h\to \tau\tau)\sim 4\times 10^{-3}\left(\frac{\gIR v}{\LIR}\right)^4\, .
\ee
This result is very close to the experimental bound, but we must in this case face   the large contribution to $\mu\to e\gamma$ discussed  above.

\section{Neutrino masses}
In this section we would like to comment on the possible origin of the neutrino masses in these scenarios.
In principle, the origin of  neutrino masses could be the same as the one discussed above 
for the other fermions, if  right-handed  neutrinos are introduced in the SM. 
Nevertheless, a simpler option   is to assume that  lepton number is  broken    at some  UV scale $\Lambda_\nu$ by higher-dimensional operators:
\begin{equation}
\frac{1}{\Lambda_{\nu}^{2d_H-1}} \bar L^c {\cal O}_H {\cal O}_H  L\,,
\label{neut}
\end{equation}
where $L$ generically denotes  a left-handed lepton.
\eq{neut}  leads to neutrino masses of order
\be
m_\nu\simeq\frac{\gIR^2v^2}{\LIR} \left(\frac{\LIR}{\Lambda_\nu}\right)^{2d_H-1} \,.
\label{neutrino}
\ee
For  $d_H =2$, $\gIR\sim 4\pi$ and $\LIR\sim 3$ TeV,
 \eq{neutrino} gives 
 \be
 m_\nu\sim 0.1-0.01\ {\rm eV}\ \ {\rm for}\ \ \   \Lambda_\nu \sim 0.8-1.5 \times 10^8\ {\rm GeV}\,. 
 \ee
  This scale $\Lambda_\nu$ 
 could be related to the scale at which  other fermion  masses are generated,  for example,
to  $\Lambda_s$ or  $\Lambda_d$.  On the other hand, large mixing angles  in the neutrino sector between two families
can be easy  obtained by requiring  the corresponding neutrino masses  to be generated at the same scale $\Lambda_\nu$.

\section{Conclusions}
 
In this work we have proposed a new realization of the flavor structure in composite Higgs scenarios. The new construction is
based on a departure from the usual partial compositeness framework for the light (i.e.~not the top quark) SM fermions,
both in the quark and lepton sector.
The main idea is to assume that the light SM fermions get their mass through effective interactions
involving fermion bilinears, namely operators of the form $\bar f_L {\cal O}_H f_R$, where ${\cal O}_H$ is a composite operator associated with the Higgs field. These Yukawa-like operators for the various fermion species are generated at
hierarchically different energy scales $\Lambda_f$, thus effectively giving rise to the hierarchy of SM fermion masses and to the structure of the CKM matrix.

The only field that does not follow this construction is the top quark, whose large Yukawa coupling
points towards a partial-compositeness origin at $\LIR\sim$ TeV, the scale at which the Higgs emerges as a composite state.  
The left-handed and right-handed top components
are thus linearly mixed with suitable composite operators, $\epsilon_{f_i} \bar f_i {\cal O}_{f_i}$,
 following the usual anarchic flavor structure.

The new framework leads to a significant improvement of the compatibility of the composite Higgs models with the flavor
constraints. The most remarkable difference with respect to the anarchic scenarios
is the  suppression of new-physics effects in dipole operators.
 The most severe bounds of the anarchic scenario, namely the ones
coming from the neutron and electron EDMs and from $\mu \rightarrow e \gamma$, are absent in
the new framework (see Table~\ref{tab:bounds2}).

The most important contributions  in our scenario come from two flavor-violating operators
arising  from the top partial compositeness.
Up to an unknown coefficient expected to be of order one,  these are given by
\be
\frac{1}{\LIR^2} \left( g_{ij}\, \bar d_{Li}  \gamma^\mu   d_{Lj}\right)^2
\ ,\ \ \ \
\frac{\gIR  v^2}{\LIR^2} g_{ij} \left(\bar d_{Li} \gamma^\mu   d_{Lj}\right)
\frac{gZ_\mu}{\cos{\theta_W}}\,, 
\label{maineffects}
\ee
where
\be
g_{ij}\equiv Y_tx_t(V^\dagger_{\rm CKM})_{i3}(V_{\rm CKM})_{3j}\,,
\ee
and $d_{Li}$ denotes the left-handed down-type quark component in the $i$-th family.
A remarkable feature of these corrections is the fact that they automatically follow a MFV structure.
The first operator  contributes to $\Delta F = 2$ transitions 
and  generates correlated effects in the $\epsilon_K$, $\Delta M_{B_d}$ and $\Delta M_{B_s}$ observables,
which are of the order of the present experimental sensitivity if we take  $\LIR\sim$ TeV and 
we allow for a  slight reduction of the left-handed top compositeness, $x_t < 1$.
The second operator of  \eq{maineffects} gives flavor-changing  $Z$-couplings. 
At present it  only  pushes the $\LIR$ scale in the $few$~TeV range. 
In the future it can be seen  either in deviations in the decays $K\to \mu\mu$ or $B\to (X) \ell\ell$.
This contribution can  however be significantly smaller
if the  strong sector  is invariant under a custodial $P_{LR}$ symmetry, which protects the down-type quark couplings
to the $Z$ boson~\cite{Agashe:2006at}.

Additional contributions to $\Delta F = 2$ operators can also be generated at the scales $\Lambda_{c,s,d}$
at which the second and first family quarks get their masses. 
These corrections however only give a sizable effect on $\epsilon_K$    for $\LIR$ below the multi-TeV range,  
a much  smaller contribution  than the anarchic one.
It must however  be stressed that these bounds depend  on the coefficients of the
effective operators which are affected by some degree of uncertainty.
These  contributions to $\epsilon_K$ severely constrain the maximal dimension of the
${\cal O}_H$ operator,  requiring $d_H \lesssim 2$.

\begin{figure}[t]
\centering
\includegraphics[width=0.82\textwidth]{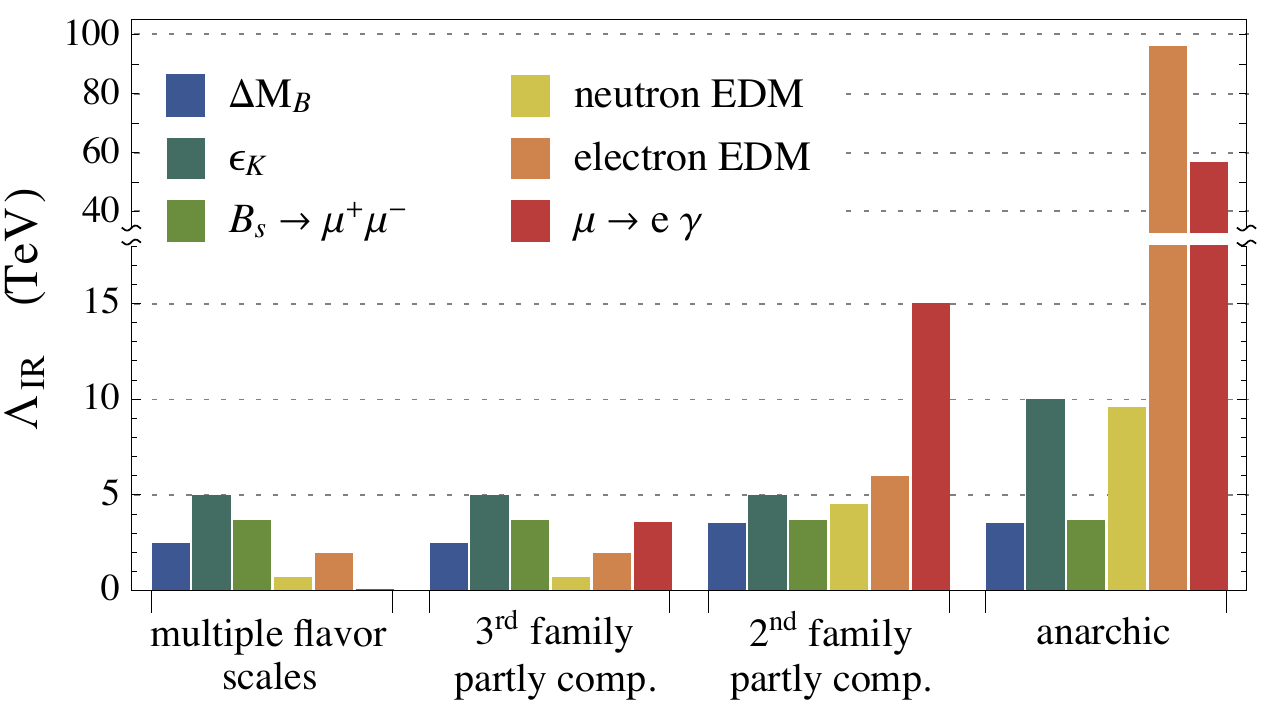}
\caption{\it Lower bounds on $\LIR$ on the various flavor scenarios. The first set of bounds corresponds to our  scenario 
with multiple flavor scales, the second and third sets assume partial compositeness at $\LIR$ 
for the whole third  and second  family respectively,
while the last set gives the bounds for the anarchic flavor scenario.
To derive the numerical values we have taken  $g_* \simeq 3$, $x_{t}\simeq x_{c} \simeq 0.5$,
and set all free  $\alpha_{L,R}$ parameters to one.}
\label{fig:bounds}
\end{figure}

We also considered possible variations of the framework described above. For example, a more  economical scenario
has  been proposed in which  each  family is associated to a single   flavor scale
 at which the bilinear mass operators are generated.
A  few additional new-physics flavor effects are
generated in this case, which are of the same order of the experimental bounds. 
In particular, assuming $\tau$ partial compositeness at $\LIR$  (as the top and bottom) leads to
corrections to the electron EDM and to the lepton-number violating processes $\mu \rightarrow e \gamma$ and
$\tau \rightarrow \mu \gamma$ which could be visible in  forthcoming experiments.
On the other hand, 
reducing down to $\LIR$    the scale at which the Yukawa interactions are generated
for the   second family
seems disfavored, since it leads to large corrections
to the neutron and electron EDMs as well as to $\mu \rightarrow e \gamma$.

Finally, we have also presented the size of  deviations in Higgs couplings, \eq{hcoup}, 
predictions for $h\to \tau\mu$, and discussed the  possible origin of the  neutrino masses.

A comparison of the bounds in the various scenarios we considered in our analysis is shown
in Fig.~\ref{fig:bounds} for a typical choice of  parameters.
We have also  included for comparison the constraints for the anarchic flavor scenario.
Fig.~\ref{fig:bounds}  shows the  main point of the article:
there are  natural  scenarios where  the origin of flavor and electroweak scale can be determined  dynamically,  
and where,  without tuning or imposing extra symmetries, 
   contributions to flavor and CP-violating observables
can still be  below (or better say, saturating) the present bounds,  providing  then a motivation for  an  experimental 
improvement  in the near future.

\medskip
\section*{Acknowledgments}
We thank Jernej  Kamenik and Andrea Wulzer for useful discussions, and especially Riccardo Rattazzi
for a critical reading of our paper.
This  work  has been partly supported by the Catalan ICREA Academia Program, and  grants FPA2014-55613-P, 2014-SGR-1450 and    Severo Ochoa excellence program  SO-2012-0234.

\appendix

\section{Warped five-dimensional   models with multiple flavor scales}
\label{adscft}

For AdS/CFT practitioners it can be useful to depict 
warped five-dimensional models which, by means of the AdS/CFT correspondence,
lead  to  the scenarios of flavor   considered above.

\begin{figure}[t]
\centering
\includegraphics[width=0.7\textwidth]{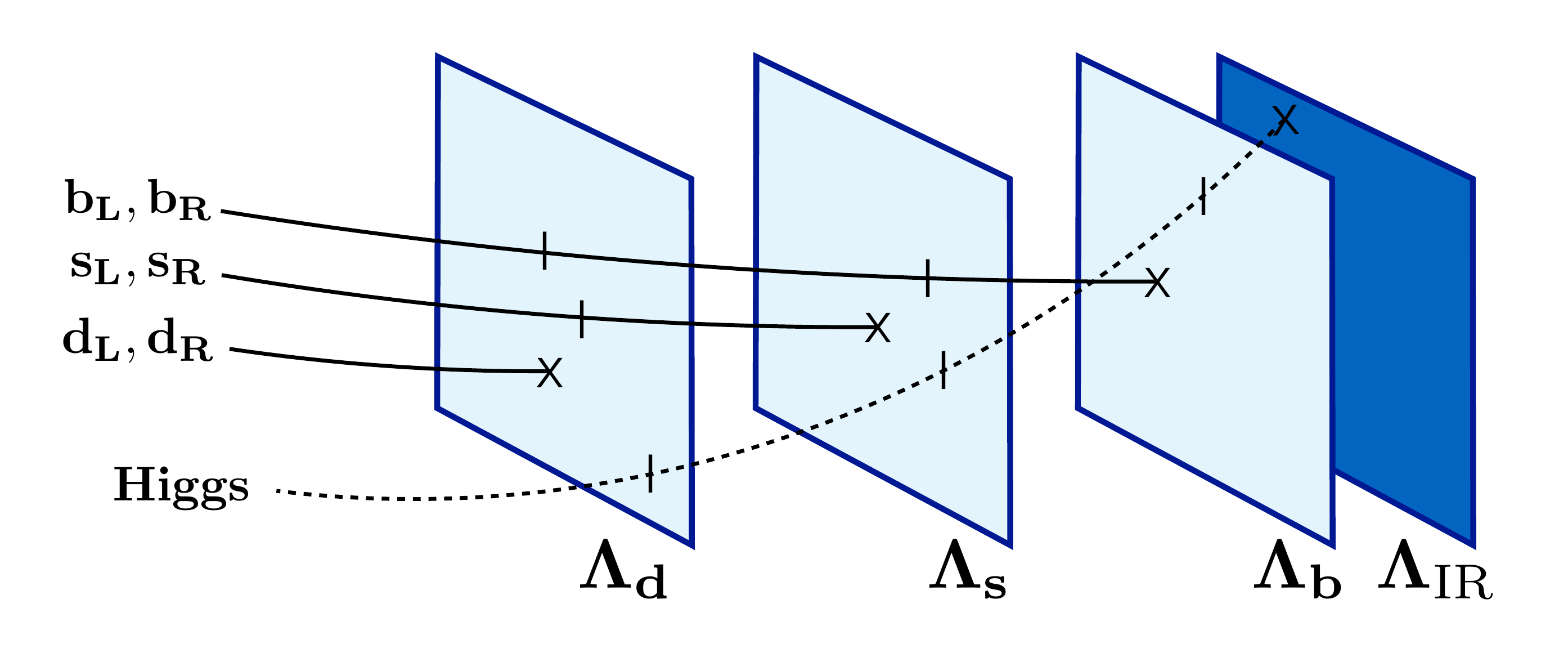}
\caption{\it Five-dimensional model  which, by  AdS/CFT, corresponds
to a model of   flavor   for the down sector and  Higgs of the SM giving  the same Yukawa structure as  \eq{yukawad}.}
\label{ads}
\end{figure}

As an example, we consider a model for the down-type quark sector and Higgs of the SM.
This is shown in Fig.~\ref{ads}.
It  corresponds to  a warped extra dimension with 
3   branes located at different  positions and therefore   associated with 3 different  energy scales 
$\Lambda_{d,s,b}$. 
We assume that only one  left-handed and right-handed quark  can propagate  up to the brane at $\Lambda_b$,
what we call the bottom quark,
while    two can propagate up to the brane at $\Lambda_{s}$.
On the other hand,  the three quarks  can be present on the brane at $\Lambda_{d}$.
 The warped extra dimension extends up to the brane at $\LIR$.
The Higgs arises from a 5D scalar field whose zero-mode is mostly localized  at $\LIR$, as shown with
a dashed line in Fig.~\ref{ads}  (the more localized towards $\LIR$, corresponds to    larger values of $d_H$).
Possible examples of    wave-functions for  the zero-modes of the quarks are also shown in Fig.~\ref{ads} with  solid lines.
Yukawa couplings come from the   overlapping of   zero-mode  wave-functions.
The small overlapping of the  Higgs wave-function  with those of the  quarks  localized   far away from $\LIR$
would   explain the smallness of these Yukawa couplings.
 The generalization to the up and lepton sector is straightforward.
If the up sector is included, one has to assure that  the left-handed doublets
  reach  also the corresponding branes where the up-type quark  masses are generated, e.g.,
   $Q_{L3}=(t_L,b_L)^T$ must reach $\Lambda_{t}$.

\section{Mixing angles in the \boldmath $\alpha_L^{ds,db,sb}\approx 0$ limit}
\label{app:z2}

Although the elements of $V_L^{\rm down}$  are fixed by the requirement of reproducing the CKM
structure, the elements of $V_R^{\rm down}$ are free parameters and could  be substantially reduced.
In this Appendix we want to show 
how small the off-diagonal entries of $V_R^{\rm down}$ could be
in the situation  where  $\alpha_L\sim 0$.

Having $\alpha_L\sim 0$ could arise from certain accidental symmetries at  $\Lambda_{d,s}$.
For example,  if at $\Lambda_d$  
there is a  $Z_2$ symmetry under which
$s_L$ and $b_L$ are odd, this would imply $\epsilon^{(1)}_{s_L,b_L}=0$.
Similarly, if  at  $\Lambda_s$
this $Z_2$ parity  is still  preserved but only for  $b_L$, we would have  $\epsilon^{(2)}_{b_L}=0$.
This would give $\alpha_L^{ds}=\alpha_L^{db}=\alpha_L^{sb}=0$.
This accidental $Z_2$ parity could arise from  the dynamics of the model.
For example, if $b_L$ is mostly composite at $\LIR$, its couplings at $\Lambda_{d,s}$ will be 
suppressed   (in  warped five-dimensional models this implies that the  wave-function of $b_L$ is peaked toward the
  $\LIR$ brane, having a small overlapping with   the $\Lambda_{d,s}$ branes). 

Having $\alpha_L\sim 0$ leads to a   $V^{\rm down}_R$  different from
\eq{rotationLR}, where $\alpha_L\sim 1$ was assumed.
For example, in the case $\alpha_L^{ds} = \alpha_L^{db} = 0$ and $\alpha_L^{sb} \sim 1$ we get
\begin{eqnarray}
(V_R^{\rm down})_{31} &\sim&\alpha_R^{ds} \alpha_L^{sb} \frac{Y_d^2}{Y_s Y_b}
\simeq (V_{\textrm{CKM}})_{21} \alpha_L^{sb} \frac{Y_d}{Y_b}
\simeq \lambda_c \alpha_L^{sb} \frac{Y_d}{Y_b}\,,\nonumber\\
(V_R^{\rm down})_{13} &\sim& \alpha_R^{db} \left(\frac{Y_d}{Y_b}\right)^2
\simeq (V_{\textrm{CKM}})_{31} \frac{Y_d}{Y_b} \simeq \lambda_c^3 \frac{Y_d}{Y_b}\,,\label{brotation}\\
(V_R^{\rm down})_{12} &\sim& (V_R^{\rm down})_{21} \sim \alpha_R^{ds} \left(\frac{Y_d}{Y_s}\right)^2
\simeq (V_{\textrm{CKM}})_{21} \frac{Y_d}{Y_s} \simeq \lambda_c \frac{Y_d}{Y_s}\,.\nonumber
\end{eqnarray}
Notice that in this case the entries involving the first and third families are not symmetric, that is 
$(V_R^{\rm down})_{13} \gg  (V_R^{\rm down})_{31}$. If $\alpha_L^{sb} = 0$ as well, the estimate for
$(V_R^{\rm down})_{31}$ coincides with the one for $(V_R^{\rm down})_{13}$, namely 
\begin{equation}
(V_R^{\rm down})_{31} \sim (V_R^{\rm down})_{13} \sim \lambda_c^3 \frac{Y_d}{Y_b}\,.
\end{equation}
Analogous results can be found for the $V_{L,R}^{\rm up}$ matrices.


\end{document}